\documentclass{iopart}

\usepackage{latexsym}
\usepackage{amssymb}
\usepackage{epsfig}

\newcommand{\connE}{\ensuremath{\tilde{\nabla}}}
\newcommand{\phib}{\ensuremath{\bar{\phi}}}

\newcommand{\nablaS}{\ensuremath{\hat{\nabla}}}

\newcommand{\metE}{\ensuremath{\tilde{g}}}
\newcommand{\metM}{\ensuremath{g}}

\newcommand{\volE}{\ensuremath{\sqrt{-\metE}}}
\newcommand{\volM}{\ensuremath{\sqrt{-\metM}}}
\newcommand{\RiemE}{\ensuremath{\tilde{R}}}

\newcommand{\EinE}{\ensuremath{\tilde{G}}}

\newcommand{\Vp}{\ensuremath{\frac{dV}{d\mu}}}

\newcommand{\grad}{\ensuremath{\vec{\nabla}}}

\newcommand{\metS}{\ensuremath{\hat{g}}}
\newcommand{\AS}{\ensuremath{\hat{A}}}

\newcommand{\KS}{\ensuremath{\hat{K}}}

\newcommand{\cS}{\ensuremath{\hat{c}}}

\begin{document}

\topical{The Tensor-Vector-Scalar theory and its cosmology}
\author{Constantinos Skordis}
\address{Perimeter Institute for Theoretical Physics, 31 Caroline street North, Waterloo, Ontario N2L 2Y5, Canada}
\ead{cskordis@perimeterinstitute.ca}

\begin{abstract}
Over the last few decades, astronomers and cosmologists have accumulated vast amounts of data clearly demonstrating that our current theories of fundamental particles and of gravity are inadequate to explain the observed discrepancy between the dynamics and the distribution of the visible matter in the Universe. 
The Modified Newtonian Dynamics (MOND) proposal aims at solving the problem by postulating that Newton's second law of motion
is modified for accelerations smaller than $\sim 10^{-10}m/s^2$.  This simple amendment, has had tremendous success in explaining galactic rotation
curves. However, being non-relativistic, it cannot make firm predictions for cosmology.

A relativistic theory called Tensor-Vector-Scalar (TeVeS) has been proposed by Bekenstein building on earlier work of Sanders which has a
MOND limit for non-relativistic systems.
 In this article I give a short introduction to TeVeS theory and focus on its predictions for cosmology as well
as some non-cosmological studies.
\end{abstract}

\pacs{98.5}
\submitto{\CQG}

\maketitle

\section{Introduction}

Over the last few decades, astronomers and cosmologists have accumulated vast amounts of data clearly demonstrating
that our current theories of fundamental particles  and of gravity are inadequate to
explain the observed  discrepancy between the dynamics and the distribution of the visible matter 
in the Universe~\cite{Oort1932,Zwicky1933,Smith1936,RubinFord1970,RubinThonnardFord1980,SofueRubin2000,CloweEtAl2006,NoltaEtAl2008,SchmidtEtAl1998,PerlmutterEtAl1998a}.
This has been called the "missing mass problem" or the "mass-discrepancy" problem.

On galactic and cosmological scales, gravity is the dominant force which drives the dynamics of all matter.
Our current well accepted theory of gravity is Einstein's General Relativity (GR) which explains how the dynamics of the various matter
species are driven by their collective energy density and pressure.
Although GR has been vigorously tested in our solar system,
 on cosmological scales and curvatures it has traditionally been assumed from the outset. Yet, it is on these scales
where the observed dynamics fail to match the observed matter distribution,
 from the scales of galaxies to the scales of the Cosmic Microwave Background (CMB) radiation. 

One could imagine that the missing mass is composed of baryons in objects other than stars,
for example Jupiter size planets or brown dwarves, collectively called MACHOS, or baryonic dark matter. 
These objects cannot 
be seen because they do not emit light of their own. However microlensing studies did not detect 
the abundance needed for these objects to make up for the missing mass~\cite{AlcockEtAl1998,AlcockEtAl2000,AfonsoEtAl2003,YooChanameGould2003}. Moreover the 
abundances of elements
predicted by Big Bang Nucleosynthesis (BBN) give a matter density far below the needed mass density~\cite{EidelmanEtAl2004,Steigman2005}.

The traditional way to explain the discrepancy between dynamics and matter distribution, is to
posit  a new form of matter, non-baryonic in nature, named  Dark Matter.
Dark Matter is thought not to interact with electromagnetic radiation
and therefore cannot be detected by observing photons at various frequencies.
Even though it cannot be seen directly, its presence is evident from the pull of gravity. Thus one attributes the extra gravitational force observed, to 
a "dark matter" component whose abundance is required to greatly exceed the visible matter abundance. 
Dark matter candidates have been traditionally split~\cite{BondSzalay1983} into "hot dark matter" and "cold dark matter", although 
an intermediate possibility, namely "warm dark matter" is sometimes being considered.

The earliest possibility considered for a dark matter candidate was a massive neutrino~\cite{GershteinZeldovich1966,MarxSzalay1976,SzalayMarx1976,CowsikMcClelland1973},
since neutrinos are particles which are known to exist as well as being very weakly interacting. 
However, massive neutrinos cannot be the dominant form of the dark matter.
If the dark matter is composed of massive neutrinos then their mass must be at 
most $30-70eV$ for reasonable values of the 
Hubble constant, if they are not to overclose the universe~\cite{GershteinZeldovich1966}.
On the other hand the Tremaine-Gunn inequality~\cite{TremaineGunn1979} gives a lower bound on the neutrino mass
if neutrinos are to be bounded gravitationally within  some radius. For example for dwarf spheroidal 
galaxies,  their mass should be greater than $\sim 300-400 eV$ which is well above the cosmologically allowed mass range.
Finally the recent Mainz and Troisk experiments
from tritium beta decay, combined with neutrino oscillation experiments give an upper limit for 
the neutrino mass  of around $2.2 - 2.5eV$~\cite{LobashevEtAl1999,WeinheimerEtAl1999,WeinheimerEtAl1999err}. 
Massive neutrinos are therefore ruled out as a dark matter candidate capable of solving the missing mass problem.

Cold dark matter (CDM), is composed of very massive slowly moving
and weakly interacting particles. A plethora of such particles generically arises in particle physics models beyond the standard model quite
naturally.  This subject~\cite{GriestKamionkowski2000,BertoneHooperSilk2004}
has been studied in great depth and has been shown to agree with observations to a very good degree.
The prospects of discovering a  CDM particle are high but so far there has not been a well accepted, firm detection.
Observationally problems persist on small scales. As simulations show, it is quite problematic to create galaxies with the right halo profile. It also seems 
 hard to explain the slope and normalization of the Tully-Fisher relation, as well as the very small scatter around it.
Moreover, CDM falls short to account for a key observation: that the expansion of the Universe is accelerating~\cite{SchmidtEtAl1998,PerlmutterEtAl1998a}.
Cosmic acceleration therefore seems to call for a new substance, collectively called the Dark Energy, which contributes a "missing energy".

The discovery that the expansion of the Universe is accelerating, gives a new twist to the whole problem, 
 since a particle like CDM cannot cause such a bizarre phenomenon. 
The Dark Energy, apparently  provides for most of the energy density in the Universe today and 
must have the very peculiar property that it provides negative pressure.
Although research has produced many proposals concerning its nature~\cite{CopelandSameTsujikawa2006}, 
there is as yet no compelling candidate for the Dark Energy. Its mere presence as well as its magnitude is
a puzzle for any sensible particle physics model. 
To quote the Dark Energy Task Force  committee report~\cite{AlbrechtEtAl2006} "nothing short of a
revolution in our understanding of fundamental physics will be required to achieve a full understanding of the
cosmic acceleration". It may well be that this  revolution requires nothing short of re-evaluating our theory of gravity.

Given that the law of gravity plays such a fundamental role at every instance where discrepancies have been observed,
it is quite possible that the phenomena commonly attributed to dark matter and dark energy are really
a different theory of gravity in disguise. This direction of 
research~\cite{Finzi1963,BekensteinMilgrom1984,KuhnKruglyak1987,Bekenstein1988,MannheimKazanas1989,Sanders1997,Drummond2001,Bekenstein2004a,Moffat2006,Krasnov2007b,KrasnovShtanov2008,Banados2008,BanadosEtAl2008}, 
 has remained largely unexplored
compared to the extensive treatment that Dark Matter (and more recently Dark Energy) has received.

Most research which concerns modifications to gravity as an explanation to
dark matter has revolved around Milgrom's Modified Newtonian Dynamics (MOND) proposal~\cite{Milgrom1983a,Milgrom1983b,Milgrom1983c}. 
Within the MOND paradigm, an acceleration scale $a_0$ is introduced. 
For accelerations smaller than $a_0$, Newton's 2nd law  is modified such that the gravitational force is proportional
to the square of the particle's acceleration. This simple amendment, has had tremendous success in explaining galactic rotation 
curves~\cite{MilgromBraun1988,BegemanSanders1991,Sanders1996,deBlokMcGaugh1998,McGaugh_deBlok1998a,MilgromSanders2004,MilgromSanders2006} 
and gives a natural explanation for the Tully-Fisher relation~\cite{McGaughEtAl2000,McGaugh2005}. 
The reader is referred to~\cite{Sanders1990,Milgrom1998c,SandersMcGaugh2002,Milgrom2008a} for reviews of MOND
and to  ~\cite{BrunetonEsposito-Farese2007} for a thorough study of
 field theoretical formulations of MOND.

To be able to make predictions for cosmological observations, a relativistic theory is needed. 
There have been a few attempts to create a relativistic theory that would encompass the MOND paradigm.
Starting with a reformulation of MOND  as a non-relativistic theory stemming from an action~\cite{BekensteinMilgrom1984}, Bekenstein and Milgrom
considered the first relativistic realization of MOND by using a scalar field~\cite{BekensteinMilgrom1984}. However, that original formulation was immediately found to
have problems with superluminal propagation.  A second attempt~\cite{Bekenstein1988} was ruled out by solar system tests.
Both of those theories were based on the existence of two metrics, related by a conformal transformation.
It was realized that the simple conformal transformation  should be changed. The reason was that
any theory based on a conformal transformation would not be able to explain the extreme bending of light  observed by massive objects from which very large
mass-to-light ratios were being inferred, without additional dark matter. As investigated by Bekenstein, one 
can change the conformal relation to a disformal one~\cite{Bekenstein1993} by including an additive tensor in
the transformation, not related to the two metrics, for example build out of the gradient of a scalar field. 
However, it was
soon recognised that any generalized scalar-tensor gravitation theory, even with a disformal relation
between the two metrics in the theory, would produce less bending
of light than GR and thus could not be used as a basis for 
relativistic MOND~\cite{BekensteinSanders1994}.

Sanders finally solved the lensing problem by introducing a unit-timelike vector field into the disformal transformation,
 in addition to the scalar field~\cite{Sanders1997}.  The vector field in Sanders's theory is however non-dynamical,
which contradicts the spirit of general covariance. 
Bekenstein generalized the Sanders theory by making the vector field dynamical, after postulating 
that its field equations stem from a Maxwellian-type action~\cite{Bekenstein2004a}. The resulting theory was called Tensor-Vector-Scalar (TeVeS)
gravitational theory (see ~\cite{Bekenstein2005a,Bekenstein2005b,Bekenstein2009a,Bekenstein2009b} for other reviews) 
and was shown in the same paper to provide for a MOND and  a Newtonian limit in the weak field non-relativistic regime,
to be devoid of acausal propagation of perturbations, to be in agreement with solar system tests and to produce the right bending of light.
 This theory is the subject of this review.

This review is organized as follows. In the section \ref{sec_fundamentals} I present the theory in its generality by discussing its dynamical 
elements and giving the actions and field equations. I also give a concise derivation of the quasistatic non-relativistic limit which leads to
MOND. In section \ref{sec_FLRW} I discuss the studies of TeVeS regarding homogeneous and isotropic cosmology.
In  section \ref{sec_linear} I move on the discussion of Linear Cosmological Perturbation Theory and focus on a striking TeVeS prediction:  that
the vector field introduced in the theory to explain gravitational lensing, is also driving cosmological structure formation.
In section \ref{sec_noncosmo} I present various non-cosmological studies of TeVeS theory, namely spherically symmetric solutions such as Black Holes and 
neutron stars, stability of spherically symmetric perturbations and gravitational collapse,
Parameterized Post-Newtonian constraints, galactic rotation curves, gravitational lensing, superluminality and the time-travel of gravitational waves.
Having described the theory and its predictions, in section \ref{sec_construction} 
I give an overview of how the theory was constructed, and give the motivation for its various elements. 
In section \ref{sec_variants} I discuss various other variants of the theory and spin-offs.
I conclude the review with an outlook, open questions and future prospects in section \ref{sec_outlook}.

\section{Fundamentals of TeVeS}
\label{sec_fundamentals}

\subsection{What are the dynamical agents?}
In General Relativity, the spacetime metric $g_{ab}$ is the sole dynamical agent of gravity. 
Scalar-Tensor theories extend this by adding a scalar field as a dynamical field mediating a spin-$0$ gravitational interaction.
TeVeS also has extra degrees of freedom,  but in addition to a scalar field $\phi$, there exists a (dual) vector field $A_a$ which also
participates into the gravitational sector.
Like GR, it obeys the Einstein equivalence principle, but unlike GR it violates the strong equivalence principle.

The original and common way to specify TeVeS theory is to write the action in a mixed frame. That is, we write the action in the 
"Bekenstein frame" for the gravitational fields, and in the physical frame, for the matter fields. In this way we ensure that the
Einstein equivalence principle is obeyed. The three gravitational fields are the metric $\metE_{ab}$ (with connection $\tilde{\nabla}_a$) 
that we refer to as the Bekenstein metric, 
the Sanders vector field $A_a$ and the scalar field $\phi$.
To ensure that the  Einstein equivalence principle is obeyed, we write the action for all matter fields, using a single physical metric 
$g_{ab}$ (with connection $\nabla_a$) that we call the "universally coupled metric"~\footnote{Some work on TeVeS, including the original articles by Sanders~\cite{Sanders1997} 
and Bekenstein~\cite{Bekenstein2004a}, refers to
the Bekenstein frame metric as the geometric metric and is denoted as $g_{ab}$, while the universally coupled metric is refered to as the physical metric
and is denoted as $\tilde{g}_{ab}$. Since it is more common to denote the metric which universally couples to matter as $g_{ab}$, in this
review we interchange the tilde. }. The universally coupled metric is algebraically defined via a disformal relation~\cite{Bekenstein1993} as
\begin{equation}
   \metM_{ab} = e^{-2\phi}\metE_{ab} - 2\sinh(2\phi)A_a A_b.
   \label{eq:metric_relation}
\end{equation}
The  vector field is further enforced to be unit-timelike with respect to the Bekenstein metric, i.e.
\begin{equation}
 \metE^{ab} A_a A_b = -1
\label{eq:A_unit}
\end{equation}
The unit time-like constraint is a phenomenological requirement for the theory to give the right bending of light (see section \ref{sec_construction}).
Using the unit time-like constraint (\ref{eq:A_unit}) it is easy to show that the disformal transformation for the inverse metric is
\begin{equation}
   \metM^{ab} = e^{2\phi}\metE^{ab} + 2\sinh(2\phi)A^a A^b
   \label{eq:inv_metric_relation}
\end{equation}
where  $A^a = \metE^{ab}A_b$.

\subsection{Action for TeVeS}
The theory is based on an action $S$, which is split as $S = S_{\metE} + S_A + S_{\phi}+S_m$, where 
$S_{\metE}$,$S_A$,$S_{\phi}$ and $S_m$ are the actions for  $\metE_{ab}$,
  vector field $A_a$, scalar field $\phi$ and matter respectively. 

As already discussed, the action for  $\metE_{ab}$, $A_a$ and $\phi$  is written using only the Bekenstein metric $\metE_{ab}$ and not $\metM_{ab}$,
and is such that $S_{\metE}$ is of Einstein-Hilbert form
\begin{equation}
   S_{\metE} = \frac{1}{16\pi G}\int d^4x \; \volE \; \RiemE,
\label{eq:S_EH}
\end{equation}
where $\metE$ and $\RiemE$ are \label{def_detE} \label{def_RE}  the determinant and scalar curvature of $\metE_{\mu\nu}$ respectively and
$G$ \label{def_Gbare} is the bare gravitational constant. The relation between $G$ and the measured Newton's constant $G_N$ will
be elaborated in a subsection \ref{sec_quasistatic}.

The action for the vector field $A_a$ is given by
\begin{equation}
	S_A = -\frac{1}{32\pi G}  \int d^4x \; \volE \; \left[ K F^{ab}F_{ab}   - 2\lambda (A_a A^a + 1)\right],
\end{equation}
where  $F_{ab} = \nabla_a A_b - \nabla_b A_a$ leads to a Maxwellian kinetic term
and $\lambda$ \label{def_lambda} is a Lagrange multiplier ensuring the unit-timelike constraint on $A_a$ and $K$
is a   dimensionless constant.  Indices on $F_{ab}$ are moved using the Bekenstein metric, i.e. $F^a_{\;\;b} = \metE^{ac} F_{cb}$.

The action for the scalar field $\phi$ is given by
\begin{equation}
    S_{\phi} = -\frac{1}{16\pi G} \int d^4x  \volE \left[ 
    \mu \; \metS^{ab}\connE_a\phi \connE_b\phi +   V(\mu) \right]
\end{equation}
 where $\mu$ \label{def_mu} is a non-dynamical dimensionless scalar field, $\metS^{ab}$ is a new metric defined by
\begin{equation}
\metS^{ab} = \metE^{ab} - A^a A^b,
\end{equation}
  and $V(\mu)$ is an arbitrary  function which typically depends on a scale $\ell_B$. Not all choices of $V(\mu)$ would give the correct Newtonian or MOND limits in a 
quasistatic situation. The allowed choices are presented in subsection \ref{sec_quasistatic}.
The metric $\metS^{ab}$ is used in the scalar field action, rather than $\metE^{ab}$ to avoid superluminal propagation
of perturbations (see section \ref{sec_construction}). It is possible to write the TeVeS action using $\metS^{ab}$, with the consequence of
having more general vector field kinetic terms (see appendix \ref{app_diagonal}).
 
The matter is coupled only to the universally coupled metric $\metM_{ab}$ and thus its action is 
of the form 
\begin{equation}
 S_m[\metM,\chi^A, \nabla \chi^A] = \int d^4x \; \volM \; L[ \metM ,\chi^A, \nabla \chi^A]
\end{equation}
for some generic collection of matter fields $\chi^A$. This defines the matter stress-energy tensor $T_{ab}$ through
 $\delta S_m = -\frac{1}{2} \int d^4x \volM \; T_{ab} \; \delta\metM^{ab}$.

\subsection{The field equations}
The field equations are easily found using the variational principle. We get two constraint equations, namely the unit-timelike constraint (\ref{eq:A_unit})
and the $\mu$-constraint
\begin{equation}
      \metS^{ab}\nabla_a\phi \nabla_b\phi  = -\Vp.
\label{eq:mu_con} 
\end{equation}
which is used to find $\mu$ as a function of $\nabla_a\phi$.

The field equations for $\metE_{ab}$ are given by
\begin{eqnarray}
   \EinE_{ab} &=& 8\pi G\left[ T_{ab} + 2(1 - e^{-4\phi})A^c T_{c(a} A_{b)}\right]
  \nonumber \\ 
&&        
+ \mu \left[ \connE_a \phi \connE_b\phi  - 2 A^c\connE_c\phi \; A_{(a}\connE_{b)}\phi  \right]
        + \frac{1}{2}\left(\mu V' -  V\right) \metE_{ab}   
  \nonumber \\ 
&&        
+   K\left[F^c_{\;\;a} F_{cb} - \frac{1}{4} F^{cd} F_{cd} \metE_{ab}\right]            - \lambda A_a A_b
\end{eqnarray}
where $\EinE_{ab}$ is the Einstein tensor \label{def_G} of $\metE_{ab}$.

The field equations for the vector field $A_a$ are
\begin{equation}
K \connE_c F^c_{\;\;a}
=  -\lambda A_a - \mu A^b\connE_b\phi \connE_a\phi  + 8\pi G  (1 - e^{-4\phi})A^b T_{ba} \label{eq:A_eq},
\end{equation}
and the field equation for the scalar field $\phi$ is 
\begin{eqnarray}
   \connE_a \left[   \mu \metS^{ab} \connE_b\phi \right] &=& 8\pi G e^{-2\phi}\left[\metM^{ab} + 2e^{-2\phi} A^a A^b\right] T_{ab}.
 \label{eq:Phi_eq}
\end{eqnarray}
The Lagrange multiplier can be solved for by contracting (\ref{eq:A_eq}) with $A^a$.

\subsection{Recovery of MOND in the quasistatic limit}
\label{sec_quasistatic}

\subsubsection{The quasistatic limit}
To recover the quasistatic limit, we impose the same assumptions as in the Parameterized Post-Newtonian (PPN) formalism~\cite{Will1981}, i.e. that
 the gravitational field is a small fluctuation about a background Minkowski spacetime and
that the matter fields can be represented by an effective perfect fluid with density $\rho$, internal energy $\Pi_E$, pressure $P$ and $3$-velocity $\vec{v}$.
 We then expand all fields in a perturbative series of successive orders in $v = |\vec{v}|$. 
As in the PPN formalism, we assume that $\frac{\partial}{\partial t} \sim O(v)$, $\Phi_P\sim \rho \sim \Pi_E \sim O(v^2)$ and $P \sim O(v^4)$, where
$\Phi_P$ is the Poisson potential constructed out of the matter density (which in the case of TeVeS is purely baryonic) and which by definition
obeys the Poisson equation $\grad^2 \Phi_P = 4\pi G_N \rho$ with Newton constant $G_N$. Finally note that to recover the
quasistatic limit we only need to keep terms up to $O(v^2)$, and we thus ignore $P$, and $\rho \Pi_E$ which are higher order. We may then take
the stress-energy tensor of matter to be that of a pressureless fluid, i.e. $T_{ab} = \rho u_a u_b$, where $u^a$ is the $4$-velocity of the fluid
normalized such that $g_{ab} u^a u^b = -1$. 

We now specify the remaining variables to $O(v^2)$. We have the universally coupled metric as
\begin{equation}
ds^2 = -(1+2\Phi_N) dt^2 + (1-2 \gamma_{PPN} \Phi_N) \delta_{ij} dx^i dx^j
\end{equation}
where $\gamma_{PPN}$ is a constant (one of the ten PPN parameters). 
The fluid velocity  is perturbed as $u^\mu = (  1 + \frac{1}{2} v^2  - \Phi_N , \vec{v})$ and $u_\mu = ( - 1 - \frac{1}{2} v^2  -\Phi_N , \vec{v})$. 
The geodesic equation for a 
point particle to $O(v^2)$, gives the non-relativistic acceleration $\vec{a}$ in terms of the potential $\Phi_N$ as
\begin{equation}
\vec{a} =  \dot{\vec{v}}
 + (\vec{v}\cdot \grad) \vec{v} 
=  - \grad \Phi_N
\label{eq_acc}
\end{equation}
which is Newton's 2nd law of motion. We thus identify $\Phi_N$ with the Newtonian potential.

We turn now to TeVeS for which we have the scalar field as $\phi = \phi_c + \varphi$ and let $\varphi \sim O(v^2)$. The Bekenstein metric is then perturbed as
\begin{equation}
d\tilde{s}^2 =  -e^{-2\phi_c}(1+2\tilde{\Phi})dt^2 +e^{2\phi_c}(\delta_{ij} + \tilde{h}_{ij}) dx^i dx^j
\end{equation}
 with $\tilde{\Phi} \sim \tilde{h}_{ij} \sim O(v^2)$. The vector field has components
 $A_\mu = e^{-\phi_c}\left(-1   -\tilde{\Phi} , \; \vec{0}\right)$ and $A^\mu =  e^{\phi_c} \left( 1    -\tilde{\Phi} , \; \vec{0} \right) $
where we have assumed that $A_i\sim O(v^3)$. This last assumption requires some further explanation. If we compare $A_a$ with $u_a$ it would seem
natural that we should assume that $A_i \sim O(v)$, however, this would also mean that either or both of $g_{0i}$ and $\metE_{0i}$ should also be of $O(v)$. 
It turns out, however, that we can always set both $A_i\sim O(v)$ and $\metE_{0i}\sim O(v)$ (and therefore also $g_{0i}$) to zero simultaneously by
performing a gauge transformation upto a curl~\footnote{
Consider first the case of GR, and assume that $g_{0i}\sim O(v)$ which leads to the field equation
\begin{equation}
\delta^{kj} \grad_i \grad_k g_{0j} - \grad^2 g_{0i}   = 0
\label{eq_GR_gauge}
\end{equation}
from the $0i$ component of the Einstein equations.  Now $g_{ab}$ is not unique but it transforms under gauge transformations generated by a vector field $\xi^a$.
 We can  choose $\dot{\xi}_0 = 0$ and $\grad_i \xi_j + \grad_j\xi_i = 0$, a choice which leaves $g_{00}$ and $g_{ij}$ invariant
 but transforms $g_{0i}$ to $g_{0i} +  \dot{\xi}_i + \grad_i \xi_0 $. Thus if we set $g_{0i}$ to zero by choosing $\xi_a$ such that
$g_{0i} = -  \dot{\xi}_i - \grad_i \xi_0$ we find that this choice also solves the Einstein equation (\ref{eq_GR_gauge}), meaning that $g_{0i}\sim O(v)$ is
pure gauge. Therefore we must have that $g_{0i}\sim O(v^3)$. 

A similar situation arises in TeVeS. If we assume that $A_i\sim O(v)$ we get the field equation
\begin{equation}
\delta^{kj} \grad_i\grad_k A_j - \grad^2 A_i  = 0
\label{eq_TEVES_gauge}
\end{equation}
However, under the special gauge transformation above we have that $A_i$ transforms to $A_i -   \grad_i \xi_0$. Furthermore, if
 $\metE_{0i}\sim O(v)$ as well, we find that it obeys a field equation like GR (\ref{eq_GR_gauge}). It is ease to show that we can simultaneously
set both the scalar mode in $A_i$ and $\metE_{0i}$ to zero by the above gauge transformation, i.e. by choosing $A_i =  \grad_i \xi_0 $ and $g_{0i}= -  \dot{\xi}_i$.
This leaves a gauge-invariant purely vector mode, i.e. $\grad_i A_i = 0$ which must therefore be given in terms of a curl as $\vec{A} = \grad \times \vec{H}$.
This curl would vanish in spherically symmetric situations and therefore does not have anything to do with the Newtonian or MOND limits. We
therefore ignore it and set $A_i\sim \metE_{0i} \sim g_{0i} \sim O(v^3)$. 
}. Therefore we set  $A_i\sim \metE_{0i} \sim g_{0i} \sim O(v^3)$. 
As the field equations also contain the functions $f(X)$ and $\mu = \frac{df}{dX}$ (see
appendix \ref{app_aqual_phi}) or equivalently $\mu$, $V(\mu)$ and $\frac{dV}{d\mu}$, we need to impose the order in $v$ for these as well.
It is clear from (\ref{eq_X}) and (\ref{eq:mu_con}) that $X\sim \frac{dV}{d\mu} \sim O(v^4)$. Furthermore, in the Newtonian limit (see further below)
we have that $f(X)\rightarrow X$, therefore $f\sim V\sim O(v^4)$. Finally on dimensional grounds we impose $\mu=\frac{df}{dX} \sim O(0)$.

We now proceed to find the quasistatic limit. 
First we find the $(ij)$-Einstein equation to $O(v^2)$ which gives $\tilde{G}_{ij} = 0$. By using the disformal transformation to get
$ \tilde{h}_{ij} =2[\varphi -\gamma_{PPN} \Phi_N  ] \delta_{ij}$ we find that $\gamma_{PPN} = 1$.
Then we use the vector field equations to solve for the Lagrange multiplier and insert it to the $(00)$-Einstein equation to $O(v^2)$ to get
\begin{eqnarray}
 \grad^2 \tilde{\Phi} = \frac{8\pi G}{2-K} \rho. 
\label{eq_Phi_tilde}
\end{eqnarray}
Finally we use the  scalar field equation which to $O(v^2)$ gives
\begin{eqnarray}
  \grad \cdot \left[  \mu \grad\varphi \right] &=& 8\pi G  \rho,
\label{eq_varphi_aqual}
\end{eqnarray}
while the Newtonian potential, related to the acceleration of particles (\ref{eq_acc}), is given via the disformal transformation as $\Phi_N = \tilde{\Phi} +\varphi$.
The above system of equations, (\ref{eq_Phi_tilde}) and (\ref{eq_varphi_aqual}) can be solved for any quasistatic situation, regardless of the boundary
conditions or the symmetry of the system in question provided a function $\mu (|\grad\varphi|)$ is supplied.

\subsubsection{Recovery of Newtonian and MOND limits}
So far the system of  equations (\ref{eq_Phi_tilde}) and (\ref{eq_varphi_aqual}) is more general than MOND. Indeed it need not even have any relation to
MOND (for example by chosing $f = X$). However, by considering the two limiting cases of Newtonian gravity for  large $|\Phi_N|$ and MOND for small $|\Phi_N|$
we can impose further constaints on the form of $f(X)$. The starting point is to combine  (\ref{eq_Phi_tilde}) and (\ref{eq_varphi_aqual}) to find a relation
between $\Phi_N$ and $\varphi$. 
 This is readily given up to a curl of a vector $\vec{S}$ as $(2-K) \grad \Phi_N =  [ 2-K + \mu] \grad\varphi  + \grad \times \vec{S}$. 
Thus we arrive at the "AQUAL equation" (see section \ref{sec_construction}) for $\Phi_N$, namely
\begin{eqnarray}
\grad \cdot \left[\mu_m \grad \Phi_N \right] =   4\pi G_N \rho
\end{eqnarray}
where $\mu_m$ is given by
\begin{eqnarray}
\mu_m =  \frac{G_N}{2G} \frac{\mu}{1  + \frac{\mu}{2-K}}
\end{eqnarray}
The ratio $G_N/G$ is not free but is found by taking the limit $\mu_m \rightarrow 1$, i.e. the Newtonian limit. Consistency then requires that $\mu \rightarrow \mu_0$
which is a constant~\footnote{The constant $\mu_0$ is related to the constant $k$ intoduced by Bekenstein as $\mu_0 = \frac{8\pi}{k}$. } contained 
in the function $f$ (or $V$) and we get the relation~\cite{BourliotEtAl2006,SagiBekenstein2008}
\begin{eqnarray}
\frac{G_N}{G} = \frac{2}{\mu_0}  + \frac{2}{2-K}.
\label{eq_G_GN}
\end{eqnarray}
The MOND limit is now clearly recovered as $\mu_m \rightarrow \frac{|\grad\Phi_N|}{a_0}$. However, due to the presence of the curl, we cannot easily
relate $\Phi_N$ to $\varphi$ in general, unless we impose the additional assumption of spherical symmetry for which the curl vanishes. In this case
of spherical symmetry,  the MOND limit gives
\begin{eqnarray}
\mu \rightarrow   \frac{2G}{G_N}  \frac{|\grad \varphi|}{a_0} = 
\frac{2G}{G_N} \frac{1}{\ell_B a_0} e^{\phi_c}  \sqrt{X}
\end{eqnarray}
where to remind the reader $X = \ell_B^2 \hat{g}^{ab} \nabla_a \phi \nabla_b \phi$. Since $\mu = \frac{df}{dX}$ we may integrate the above equation to
find the function $f(X)$ which in the MOND limit should be
\begin{eqnarray}
 f \rightarrow \frac{2}{3\ell_B a_0} \frac{1}{ \frac{1}{\mu_0}  + \frac{1}{2-K}} e^{\phi_c}   X^{3/2} 
\end{eqnarray}
where the integration constant has been set to zero, as it can always be absorbed into a cosmological constant for the metric $\metE_{ab}$.
Since both $X$ and $f$ are dimensionless we may define a new constant $\beta_0$ such that $a_0$ is a derived quantity given by
\begin{eqnarray}
a_0  = \frac{2}{3\beta_0\ell_B} \frac{1}{ \frac{1}{\mu_0}  + \frac{1}{2-K}} e^{\phi_c}   
\label{eq_def_a_0}
\end{eqnarray}
and the function has the MOND limit $f \rightarrow \beta_0 X^{3/2}$.
Since in the Newtonian limit we also have $f\rightarrow \mu_0 X$,  there are at least three constants that can appear in $f(X)$, namely $\mu_0$, $\beta_0$
and $\ell_B$.  

\subsubsection{Remark :}
In terms of the function $\frac{dV}{d\mu}$ the  MOND limiting case implies that $\frac{dV}{d\mu}\rightarrow  -\frac{4}{9\beta_0^2\ell_B^2} \mu^2$ 
as $\mu\rightarrow 0$ while it must diverge as $\mu\rightarrow \mu_0$ in the Newtonian limit. 
This second limit could be imposed if $\frac{dV}{d\mu}\rightarrow (\mu_0 - \mu)^{-m}$
for some power $m$. Bekenstein choses this to be $m=1$ although other choices are equally valid, even functions that have essential singularities.

\subsubsection{Remark :}
It is clear from (\ref{eq_def_a_0}) that  $a_0$  depends on the cosmological boundary condition $\phi_c$ which can defer
for each system depending on when the system in question was formed. It could thus be considered as a slowly varying function of time. 
This possibility has been investigated by Bekenstein and Sagi~\cite{BekensteinSagi2008} and by Limbach, Psaltis and Feryal~\cite{LimbachPsaltisFeryal2008}.

\subsubsection{Remark :}
The two limiting cases for $f(X)$ are somewhat strange. In particular we require that $f(X)\rightarrow X$ for $X\gg 1$ to recover the Newtonian
limit, and that  $f(X)\rightarrow X^{3/2}$ for $X\ll 1$ , in other words a higher power, to  recover the MOND limit.
This signifies that in these kind of formulation of relativistic MOND, i.e. in terms of a scalar field, the function $f(X)$ should be non-analytic. 
It further signifies that $f(X)$ can be expanded in positive powers of $\sqrt{X}$ for small $X$ and in positive powers of $\frac{1}{X}$ 
for large $X$ but  the two expansions
cannot be connected. In otherwords it is impossible to perturbatively connect the Newtonian with the MOND regime via a perturbation series in $|\grad \varphi|$.

\subsubsection{Remark :}
The Bekenstein free function in~\cite{Bekenstein2004a} is given in the notation given in this review as
\begin{equation}
  \frac{dV}{d\mu} = -\frac{3}{32\pi \ell_B^2 \mu_0^2}\frac{\mu^2 (\mu-2\mu_0)^2}{\mu_0-\mu} 
\label{eq_beke_fn}
\end{equation}
which means that $\beta_0 =  \frac{4}{3} \sqrt{\frac{2\pi\mu_0}{3}} $ and thus 
\begin{eqnarray}
a_0  = \frac{\sqrt{3}}{ 2 \sqrt{2\pi\mu_0}\ell_B} \frac{1}{ \frac{1}{\mu_0}  + \frac{1}{2-K}} e^{\phi_c}   
\label{eq_a_0_bek}
\end{eqnarray}
This is in agreement with~\cite{BekensteinSagi2008} (the authors of~\cite{BourliotEtAl2006} have erroneously inverted a fraction in their definition of $a_0$).

\subsubsection{Remark :}
One can find constraints on the form of $f(X)$ by considering the first order departures from the Newtonian limit~\cite{Bekenstein2004a}.
Based on the results found by Bekenstein~\cite{Bekenstein2004a} we can assume that $\mu = \mu_0   + \mu_1\epsilon + \frac{1}{2}\mu_2\epsilon^2 + \ldots$
where $\epsilon = 1/(\ell_B |\grad\varphi|)$. The series coefficients $\mu_i$ depend on the choice of the function $f$ and for the case of Bekenstein's function in
~\cite{Bekenstein2004a} are $\mu_1=0$ and $\mu_2 = -\frac{\mu_0^4 a_0^2 \ell_B^2}{32} $. 
These coefficients can in principle be used to put constraints on the parameters of the free function. Bekenstein notes that for $\mu_0 > 830$ the
deviation of $\grad\Phi_N$ from the Newtonian value at the orbit of the Earth is $<5.3 \times 10^{-9}$. Further out, to the orbits of the giant planets,
it may lead to the deviations recently named the "Pioneer anomaly"~\cite{TuryshevNietoAnderson2005}.

The use of solar system constraints however should be reworked as the solar system is actually in the external field of the galaxy.
It is well known that external field effects are important in MOND and the same should hold for TeVeS.

\subsection{Where TeVeS can be different than MOND}
It is evident from the discussion of the quasistatic limit that exact MOND behaviour is recovered for exact spherical symmetry.
For asymmetric situations, the presence of the curl forbids the formulation of the system of equations (\ref{eq_Phi_tilde}) and (\ref{eq_varphi_aqual}) 
in terms of a single AQUAL equation. One should therefore (at least in principle) solve (\ref{eq_Phi_tilde}) and (\ref{eq_varphi_aqual}) on its own,
after imposing the Newtonian and MOND limit on the form of $f(X)$ and thus $\mu(X)$. 

As Bekenstein argues~\cite{Bekenstein2004a}, however, the curl $\grad \times \vec{S}$ can in most situations be neglected, so that an effective MOND 
law is recovered even in asymmetric systems. Thus the curl should be important in the inner regions or near-exteriors
of galaxies but not far away from them. As Bekenstein points out~\cite{Bekenstein2004a} if a system 
is asymmetric but very dense so that the Newtonian regime applies everywhere, it is quite safe to neglect the curl.
Note also that in principle it is also possible to have an $O(v)$ curl coming from the vector field, i.e. $\vec{A} = \grad\times \vec{H} \sim O(v)$.
If this is the case, then the $O(v^2)$ equations would have terms involving $|\vec{A}|^2$. This would also give corrections to the MOND (and also to the Newtonian) limit
for aspherical systems and since this curl is not sourced by matter, Bekenstein's argument need not apply. Its implications should therefore be investigated further.

We also note that the recovery of MOND is done only in the quasistatic limit. In all situations where this limit does not apply (such as cosmology),
we should not expect to get any relation to MOND, at least not in TeVeS theory. Finally, departures from MOND behaviour are to be expected in cases
where one must have the equations to $O(v^3)$ or $O(v^4)$, or where the cosmological background
should be included as a time-dependent background (which may change the order of some terms).
 This possibility may include galaxy clusters and deserves further inverstigation. Could the vector field which plays a role for
driving large scale structure formation (see below) reconcile the problems of MOND with clusters?

\section{Homogeneous and Isotropic cosmology}
\label{sec_FLRW}

\subsection{FLRW equations}
Solutions to the TeVeS equations for a homogeneous and isotropic spacetime described by a Friedmann-Lemaitre-Robertson-Walker (FLRW) metric
have been extensively studied~\cite{Bekenstein2004a,HaoAkhoury2005,SkordisEtAl2006,Diaz-RiveraSamushiaRatra2006,DodelsonLiguori2006,Skordis2006,BourliotEtAl2006,Zhao2006a,Skordis2008a,FerreiraSkordisZunkel2008}.

For an FLRW spacetime, the universally coupled metric takes the conventional synchronous form
\begin{equation}
 ds^2 = -dt^2 + a^2 \delta_{ij}dx^i dx^j
\end{equation}
for physical scale factor $a$ and where I have assumed for simplicity that the spatial hypersurfaces are flat (see ~\cite{Skordis2006} for the curved case).
The Bekenstein metric also has a similar form, i.e. it is given as
\begin{equation}
 d\tilde{s}^2 = -d\tilde{t}^2 + b^2 \delta_{ij}dx^i dx^j
\end{equation}
for a second scale factor $b$.  The disformal transformation relates the two scale factors as $a = b e^{-\phi}$
while the two coordinate times $t$ and $\tilde{t}$ are related by $dt = e^{\phi}d\tilde{t}$.
The physical Hubble parameter is defined as usual as
\begin{equation}
 H = \frac{d\ln a}{dt} =  \frac{\dot{a}}{a}
\end{equation}
while the Bekenstein frame Hubble parameter is $\tilde{H} = \frac{d\ln b}{d\tilde{t}} =  \frac{b'}{b} = e^\phi H + \phi'$.
Cosmological evolution is governed by the Friedmann equation
\begin{equation}
3\tilde{H}^2 = 8\pi G e^{-2\phi} \left( \rho_\phi + \rho\right)
\label{eq_beke_friedmann}
\end{equation}
where $\rho$ is the physical matter density which obeys the energy conservation equation with respect to the universally coupled metric 
and where the scalar field energy density is
\begin{equation}
\rho_\phi = \frac{e^{2\phi}}{16\pi G}\left( \mu \Vp + V \right) 
\end{equation}
Similarly one can define a scalar field pressure as
\begin{equation}
P_\phi = \frac{e^{2\phi}}{16\pi G}\left( \mu \Vp - V \right) 
\end{equation}
The scalar field evolves according to the two differential equations
\begin{equation}
 \phi' = -\frac{1}{2\mu}\Gamma
\end{equation}
and
\begin{equation}
 \Gamma' + 3 \tilde{H} \Gamma = 8\pi G e^{-2\phib} (\rho + 3 P)
\end{equation}
while $\mu$ is found by inverting ${\phi'}^2 = \frac{1}{2}\frac{dV}{d\mu}$

It is important to note that the vector field must point to the time direction and its components are given by $A_{\mu} = (\sqrt{\metE_{00}},\vec{0}) $.
Therefore  it does not contain any independent dynamical information and it does not  explicitly contribute to the energy density. Its only effect is on
the disformal transformation which relates the Bekenstein-frame Friedmann equation (\ref{eq_beke_friedmann}) with the physical Friedmann equation.
This is true also in cases where the vector field action is generalized and where the only effect is a constant rescaling of the left-hand-side
of the  Bekenstein-frame Friedmann equation (\ref{eq_beke_friedmann}) as discussed by Skordis in ~\cite{Skordis2008a}.

\subsection{The Bekenstein function}
In the original TeVeS paper~\cite{Bekenstein2004a} Bekenstein studied the cosmological evolution of an FLRW universe in TeVeS
assuming that the free function is given by (\ref{eq_beke_fn}).
He showed that the scalar field contribution to the Friedmann equation is very small, and that $\phi$ evolves very little from the early universe
until today. He noted that with this choice of function, a cosmological constant term has to be added in order to have an accelerated expansion today
as required by the SN1a data.

Many cosmological studies, for example~\cite{HaoAkhoury2005,SkordisEtAl2006,DodelsonLiguori2006,Skordis2008a}, have used this Bekenstein function. 
In particular Hao and Akhoury~\cite{HaoAkhoury2005} noted that the integration constant obtained by integrating (\ref{eq_beke_fn}) can be
used to get an accelerated expansion and noted that TeVeS has the potential to act as dark energy. However, such an integration constant cannot
be distinguished from a bare cosmological constant term in the Bekenstein frame. Thus it is dubious whether this can be interpreted as 
dark energy arising from TeVeS. Nevertheless, it would not be a surprizing result if some other TeVeS functions could in fact provide for dark energy,
because of the close resemblance of the 
scalar field action in TeVeS and k-essence~\cite{Armendariz-PiconDamourMukhanov1999,Armendariz-PiconMukhanovSteinhardt2000}.
Indeed Zhao~\cite{Zhao2006a} has investigated this issue further (see below).

\begin{figure}
\epsfig{file=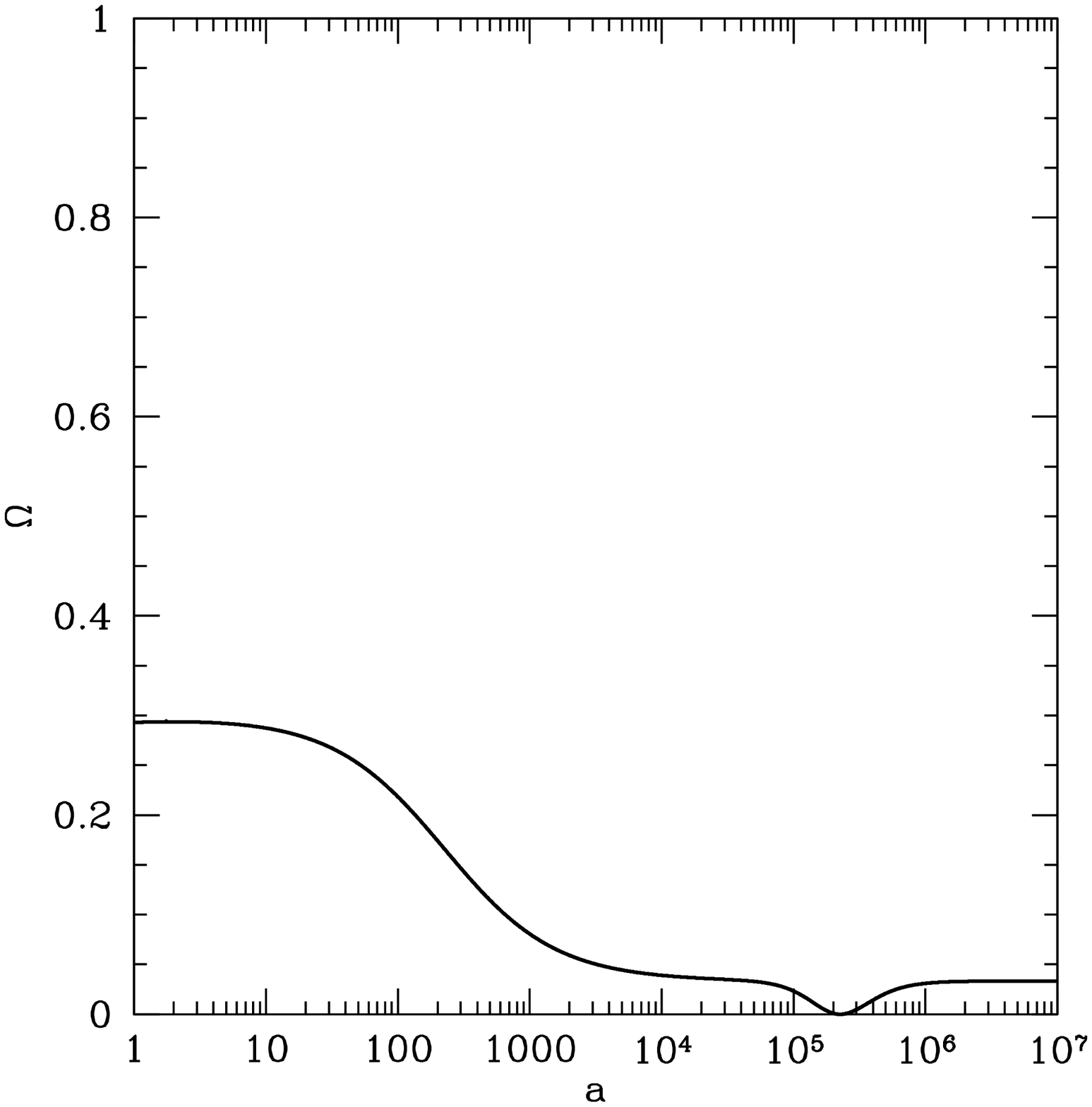,width=6.5cm}
\epsfig{file=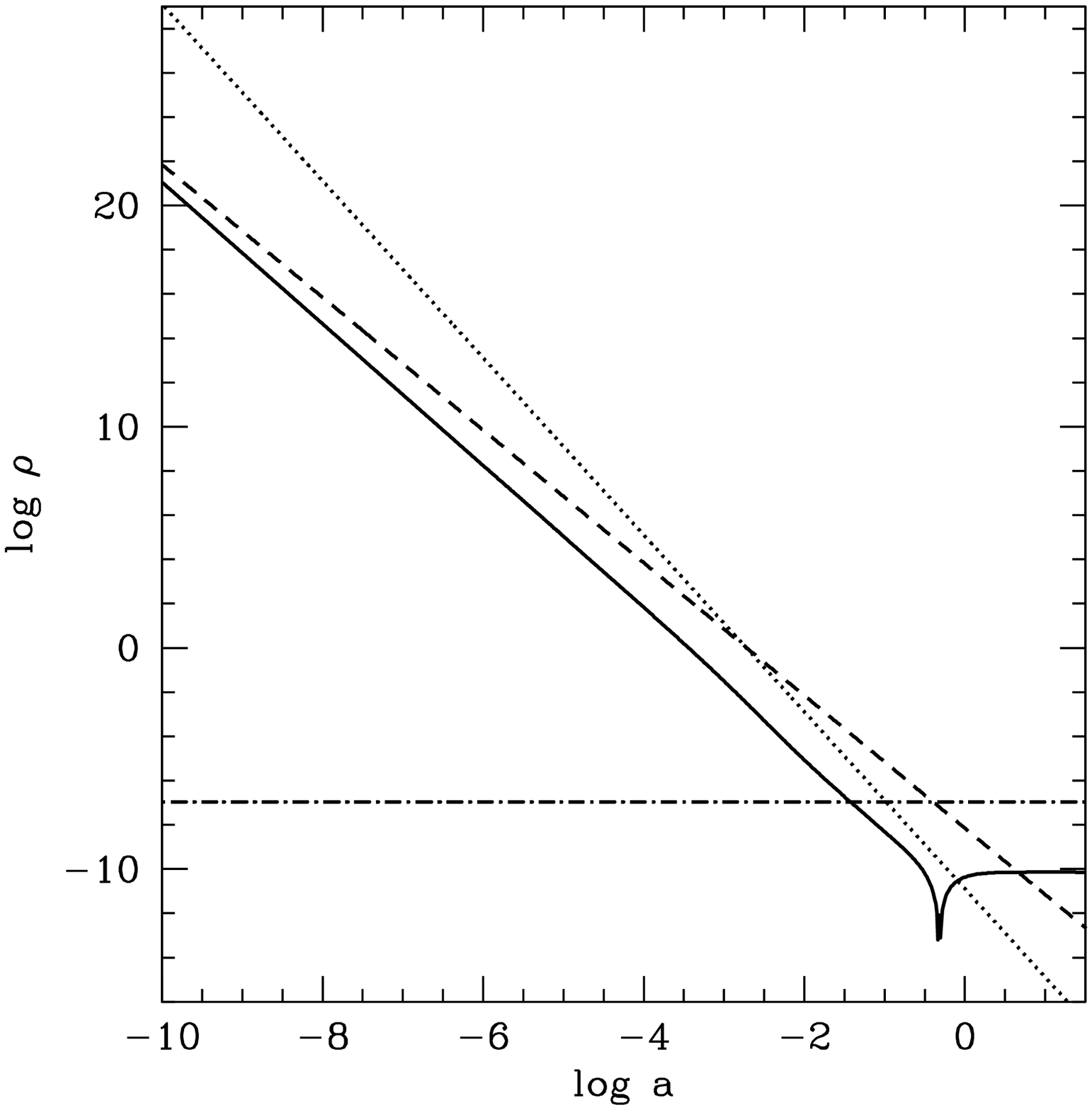,width=6.5cm}
\caption{
LEFT : The cosmological evolution of the relative density of the scalar field $\Omega_\phi$ with $\mu_0=5$ in a universe with radiation, baryons and 
a cosmologicical constant. The energy densities of the matter components have been adjusted to unrealistic values to show the tracking behavior of the
scalar field clearly.
\newline
RIGHT : The cosmological evolution of the energy densitiy $\rho$ with the scale factor $a$ for baryons (dashed), radiation (dotted), cosmological constant (dot-dash)
and scalar field $\phi$  (solid). Here $\mu_0=250$ and the energy densities of the matter components correspond to realistic values typical in
our universe.
\label{fig_omegas}
 }
\end{figure}

Exact analytical and numerical solutions with the Bekenstein free function (\ref{eq_beke_fn}) have been found by Skordis \etal~\cite{SkordisEtAl2006}
and by Dodelson and Liguori~\cite{DodelsonLiguori2006}. It turns out that not only, as Bekenstein noted, the scalar field is subdominant, but also its 
energy density tracks the matter fluid energy density.  This kind of tracker behaviour has been found before in scalar field models of dark
energy~\cite{Wetterich1994,FerreiraJoyce1997a,FerreiraJoyce1997b,AlbrechtSkordis1999,BarreiroCopelandNunes1999,Amendola2000,SkordisAlbrecht2000,Tocchini-ValentiniAmendola2001} called tracking quintessence. 
The ratio of the energy density of the scalar field to that of ordinary matter is approximately constant (see left panel of Figure \ref{fig_omegas}), so that the scalar field  exactly tracks the matter dynamics.
In the case of TeVeS, one gets that 
\begin{equation}
\Omega_\phi = \frac{(1+3w)^2}{6(1-w)^2\mu_0}
\end{equation}
where $w$ is the equation of state of the background fluid~\footnote{This excludes the case of a stiff fluid with $w=1$.}.
Since $\mu_0$ is required to be very large, the scalar field relative energy density is always small, with values typically smaller than $\Omega_\phi \sim 10^{-3}$
in a realistic situation. 
As investigated by Skordis~\cite{Skordis2008a} the tracker solutions are also present (for this choice of function) in versions of TeVeS with more general
vector field actions.

In realistic situations, the radiation era tracker is almost never realized, as has been noted by Dodelson and Liguori. Rather, during the radiation
era, the  scalar field energy density is subdominant but slowly growing (see right panel of Figure \ref{fig_omegas} ) and the scalar field is given by $\phi \propto a^{4/5}$.
Upon entering the matter era it settles into the
tracker solution. This transient solution in the radiation era has been generalized by Skordis~\cite{Skordis2008a} to an arbitrary initial condition for $\phi$,
a more general free function (see below) and a general vector field action. It should be stressed that the solution in the radiation era is important for
setting up initial conditions for inhomogeneous perturbations about the FLRW solutions, relevant for studying the physics of 
the CMB radiation and Large Scale Structure (LSS).

\begin{figure}
\epsfig{file=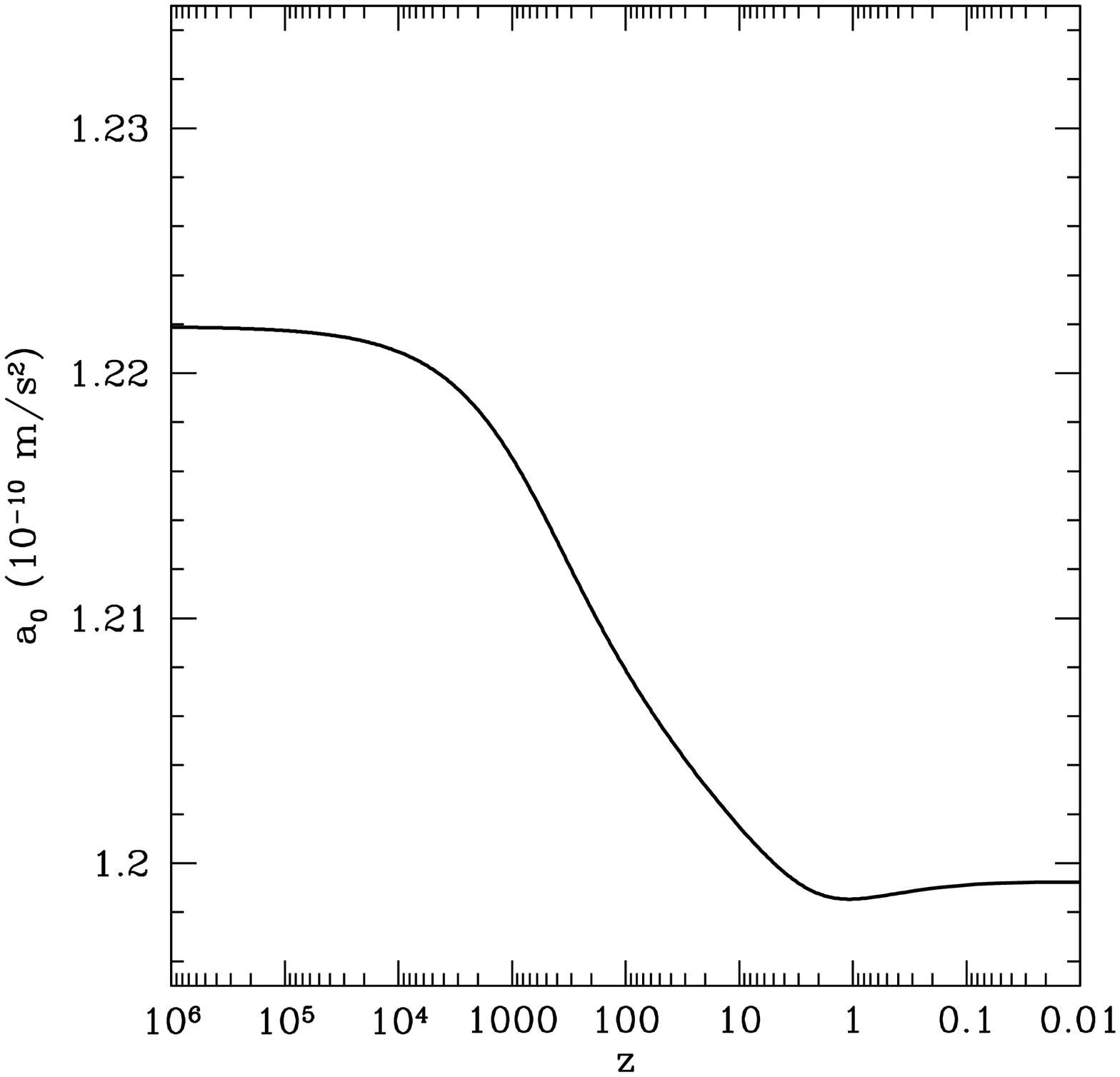,width=6.5cm}
\epsfig{file=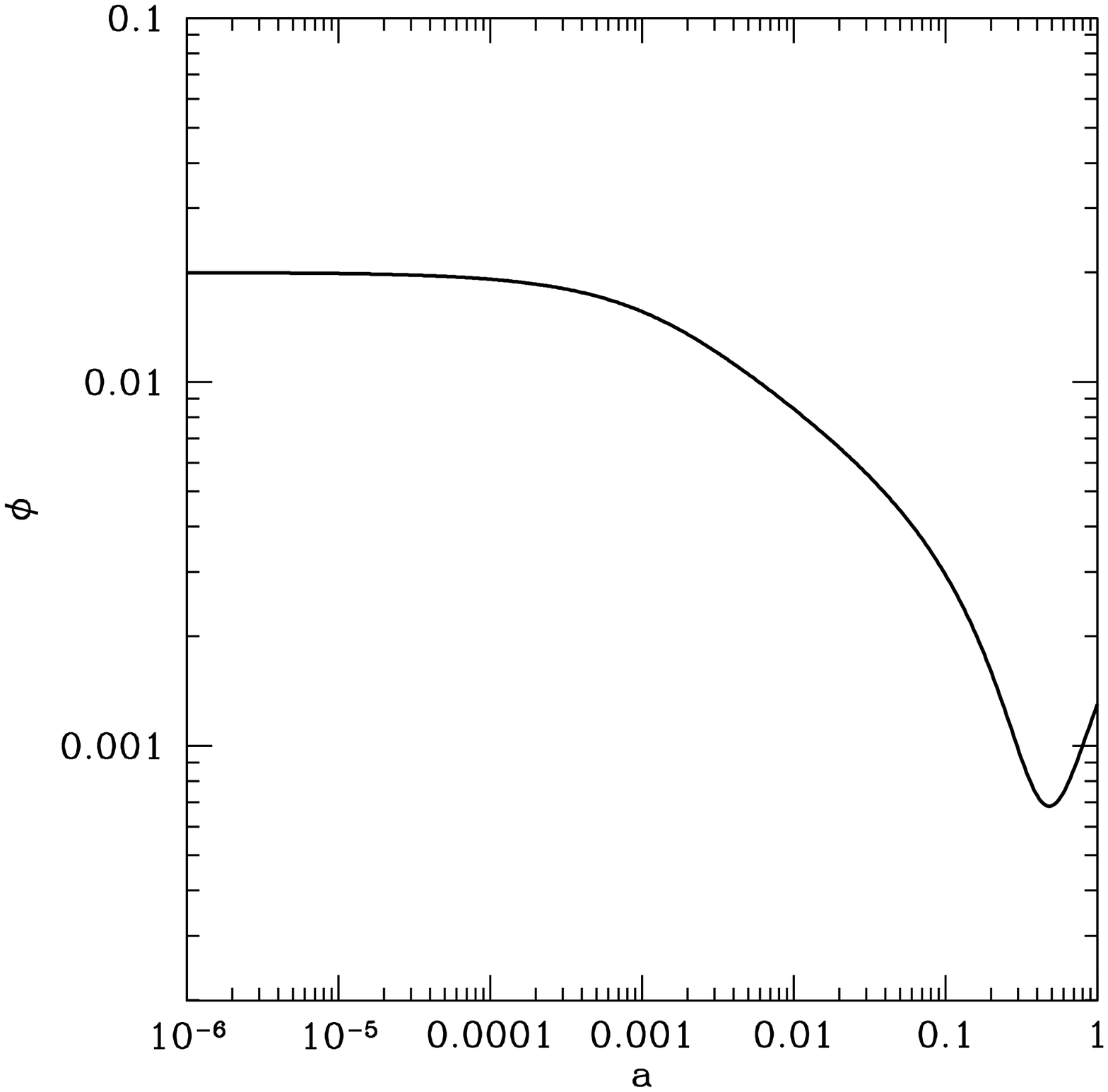,width=6.5cm}
\caption{Left : The time evolution of the Mondian acceleration $a_0$ with redshift,  for the Bekenstein function  in a typical background cosmological model.
\newline
Right : The time evolution of $\phi$ for the Bekenstein function with the scale factor for the same model.
\label{fig_a0}
 }
\end{figure}

Since $a_0$ for a quasistatic system depends on the cosmological value of the scalar field at the time the system broke off from the expansion and collapsed
to a bound structure, it is possible that different systems would exhibit different values of $a_0$ depending when they formed. Figure \ref{fig_a0}
displays the evolution of $a_0$ and the scalar field $\phi$ for realistic cosmological model. The impact of evolving $a_0$ on
 observations has been investigated by~\cite{BekensteinSagi2008,LimbachPsaltisFeryal2008}.

Finally, the sign of $\dot{\phi}$ changes between the matter and cosmological constant eras. In doing so, the energy density of the scalar field goes momentarily
through zero, since it is purely kinetic and vanishes for zero $\dot{\phi}$~\cite{SkordisEtAl2006}.

\subsection{Generalizing the Bekenstein function}
 Bourliot \etal~\cite{BourliotEtAl2006} studied more general free functions which have the Bekenstein function as a special case.
In particular they introduced two new parameters, a constant $\mu_a$ and a power index $n$ such that the free function is generalized to
\begin{equation}
\frac{dV^{(n)}}{d\mu} = 
 -\frac{3}{32\pi \ell_B^2 \mu_0^2}\frac{\mu^2 (\mu-\mu_a\mu_0)^n}{\mu_0-\mu} 
\end{equation}
This function~\footnote{Note that~\cite{BourliotEtAl2006} uses a different normalization for $V$ and their results can be
recovered by rescaling the scale $\ell_B$ in this report as $\ell_B\rightarrow\ell_B \sqrt{\frac{3}{2}\mu_0^{n-3}}$.} retains the property of having a Newtonian limit as $\mu\rightarrow \mu_0$ and a MOND limit as $\mu\rightarrow 0$. Furthermore,
as remarked in~\cite{BourliotEtAl2006}, more general functions can be built by considering the sum of the above prototypical function with 
arbitrary coefficients, i.e. as $\frac{dV}{d\mu} = \sum_n c_n \frac{dV^{(n)}}{d\mu} $.
The cosmological evolution depends on the power index $n$ and can be categorized as follows.

\subsubsection{Case $n\ge 1$}
Clearly $\frac{dV}{d\mu}(\mu_a\mu_0) = 0$ and at that point $\dot{\phi}\rightarrow 0$.
Now suppose that the integration constant is chosen such that $V(\mu_a\mu_0)=0$ as well. Then, just like the case of the Bekenstein function (which is included
in this sub-case as $n=2$ and $\mu_a=2$), one gets tracker solutions. The function $\mu$ is driven to $\mu=\mu_a\mu_0$ at which point $\dot{\phi}=0$.
There are no oscillations around that point, but it is approached slowly so that it is exactly reached only in the infinity future. The scalar field relative density
is given by
 \begin{equation}
\Omega_\phi = \frac{(1+3w)^2}{3\mu_a(1-w)^2\mu_0}
\end{equation}
 independently of the power index $n$. It should also be pointed out that the evolution of the physical Hubble parameter $H$ would be different than
the case of $GR$ even in the tracking phase~\cite{BourliotEtAl2006}. For example in the case $w=0$ we would have $H \propto a^{-n_h}$ 
where $n_h =\frac{1+3\mu_a\mu_0}{2(\mu_a\mu_0-1)}$.

Furthermore, just like the Bekenstein case, the radiation era tracker is untenable for realistic cosmological evolution for which $\mu_0$ must be large
so that $\Omega_\phi$ would be very small ($\sim <10^{-3}$). As was shown in~\cite{Skordis2008a}, 
in this case we once again get a transient solution where the scalar field evolves as $\phi \propto a^{4/(3+n)}$.

In the case that the integration constant is chosen such that $V(\mu_a\mu_0)\ne0$ then one has an effective cosmological constant present. Thus
once again we get tracker solutions until the energy density of the Universe drops to values comparable with this cosmological constant, at which case
tracking comes to an end, and the universe enters a de Sitter phase.

\subsubsection{Case $-3\le n\le0$}
The cases $n=0$ to $n=-2$ turn out to be pathological as they lead to singularities in the cosmological evolution~\cite{BourliotEtAl2006}.
The case $n=-3$ is well behaved when the matter fluid is a cosmological constant, however, it also is pathological when
the matter fluid has a different equation of state than $w=-1$~\cite{BourliotEtAl2006}.

\subsubsection{Case $n\le -4$}
The cases for which $n\le-4$ are  well behaved in the sence that no singularities occur in the cosmological evolution.
Contrary to the $n\ge 1$ cases the cosmological evolution drives the function $\mu$ to infinity. Moreover these cases do not
display the tracker solutions of $n\ge 1$, but rather the evolution of $\rho_\phi$ is such that it evolves more rapidly than the matter density $\rho$
and so quickly becomes subdominant. The General Relativistic Friedmann equation is thus recovered, i.e. $3 H^2 = 8\pi G \rho$.
We also get that $\tilde{H} = H$ which means that the scalar field is slowly rolling.

The evolution of the scalar field variables $\Gamma$, $\phi$ and $\mu$ then depends on the equation of state of the matter fluid.
If the background fluid is a cosmological constant, then we get de Sitter solutions for both metrics  and we get that $\Gamma= 2H(e^{-3Ht} -1)$.

For the case of a stiff fluid with equation of state $w=1$, we get that $\Gamma$ has power-law solutions in inverse powers of $t$ as $\Gamma = \frac{6}{t} + \frac{\Gamma_0}{t^3}$. A similar situation arises for  $-1 < w < 1$ for which we get $\Gamma = \frac{2(1+3w)}{1-w} H $  and the Hubble parameter evolves as $H  = \frac{2}{3(1+w)}\frac{1}{t}$.
Notice that the limit $w\rightarrow 1$ for the $-1<w<1$ case does not smoothly lead to the $w=1$ case.

The observational consequences on the CMB and LSS have not been investigated for this case of function, unlike
the case of the Bekenstein function.

\subsubsection{Mixed cases}
Mixing different powers of $n\ge 1$ leads once again to tracker solutions. One may have to add an integration constant in order to keep $V(\mu_a \mu_0) = 0$,
although for certain combinations of powers $n$ and coefficients $c_i$ it is not necessary.

Mixing $n=0$ with some other $n\ge 1$ cannot remove the pathological situation associated with the $n=0$ case. Mixing $n=0$ with both positive and negative powers
could however lead to acceptable cosmological evolution since the effect of the negative power is to drive $\mu$ away  from the $\mu=\mu_a \mu_0$ point.

In general if we mix an arbitrary number of positive and negative powers we would get tracker solutions provided we could expand the new function in positive
definite powers of $\mu - \mu_a'\mu_0$ where $\mu_a'$ is some number different from the old $\mu_a$.

\subsection{Inflationary/accelerated expansion solutions}
Diaz-Rivera, Samushia and Ratra~\cite{Diaz-RiveraSamushiaRatra2006} have 
studied cases where the cosmological TeVeS  equations lead to inflationary/accelerated expansion solutions.
They first consider the vacuum case, where the matter density $\rho$ vanishes. In that case, they find that one gets de Sitter solutions 
$b \sim e^{\tilde{H}_0 \tilde{t}}$ where the Bekenstein frame Hubble constant $\tilde{H}_0$ is given by the free function 
as $\tilde{H}_0 = \sqrt{\frac{\mu_0^2 V}{6}}$ and where $\frac{dV}{d\mu}=0$, i.e.
the scalar field is constant $\phi = \phi_i$. It is clear that such a solution will always occur (in vacuum) provided the free function
satisfies $\frac{dV}{d\mu}(\mu_v)=0$ and $V(\mu_v)\ne0$ for some constant $\mu_v$. In that case, the general solution is obviously not de Sitter since 
both $\phi$ and $\mu$ will be time-varying but will tend to this vacuum solution as $\mu \rightarrow \mu_v$. 
Indeed the $n\ge 1$ case of Bourliot \etal~\cite{BourliotEtAl2006} with an integration constant is precisely this kind of situation.

In the non-vacuum case, for which the matter fluid has equation of state $P = w\rho$, they make the ansatz $b^{3(1+w)} = e^{(1+3w)\phi}$
which brings the Friedmann equation into $3\tilde{H}^2 = 8\pi G \rho_0 + \frac{1}{2}(\mu V' + V)$, where $\rho_0$ is the matter density at $a=1$.
Once again they assume that the free-function-dependent general solution drives $\mu$ to a constant $\mu_v$ but $\phi$ is evolving. Thus we must have that
$\phi = \phi_1 \tilde{t} + \phi_2$, such that $\dot{\phi} = \phi_1$ is a constant. In order for $\phi_1$ to be non-zero we must have $\frac{dV}{d\mu}(\mu_v)\ne 0$.
However,  there is a drawback of this situation. As they point out, consistency with the scalar field equation requires that $w<-1$. 
Furthermore, although this solution is a de Sitter solution in the Bekenstein-frame, it corresponds to a power-law solution for the universally coupled metric.
In order for this power-law solution to lead to acceleration, they find that  $  -5/3 < w < -1$. Since no known fluids exist in this range of $w$ 
this solution is of dubious importance. 

\subsection{Accelerated expansion in TeVeS}
The simplest case of accelerated expansion in TeVeS is provided by using a cosmological constant term. This is equivalent to
adding an integration constant to $V(\mu)$~\cite{HaoAkhoury2005,BourliotEtAl2006} and it corresponds to the accelerated expansion considered by
Diaz-Rivera, Samushia and Ratra~\cite{Diaz-RiveraSamushiaRatra2006} in both the vacuum or non-vacuum cases (above). 
This kind of solution does not make things any better than a cosmological constant
model and all of its associated problems and coincidences. It would be therefore be of importance to find accelerated solutions in TeVeS without such
a constant, simply by employing the scalar field (these need not be de Sitter solutions).

Zhao used a simple function $\frac{dV}{d\mu} \propto \mu^2$ to obtain solutions which provide acceleration, and compared his solution with 
SN1a data~\cite{Zhao2006a}, finding good agreement. However, it is not clear whether other observables such as the CMB angular power 
spectrum or observations of Large Scale Structure are compatible with this function.  Furthermore, this function is certainly not realistic as it 
does not have a Newtonian limit (it is always MONDian).

Although no further studies of accelerated expansion in TeVeS have been performed, it is very plausible that certain choices of function will
inevitable lead to acceleration. It is easy to see that the scalar field action has the same form 
as a k-essence/k-inflation~\cite{Armendariz-PiconDamourMukhanov1999,Armendariz-PiconMukhanovSteinhardt2000} action which has been considered as
a candidate theory for acceleration.
 More precisely, the Friedmann-fluid-scalar field system of cosmological equations corresponds to k-essence
coupled to matter. It is unknown in general whether this has similar features as the uncoupled k-essence, although  Zhao's study
indicates that this a promising research direction. 
Let us also note that disformal transformations can also play a crucial role in theories of acceleration even for canonical scalar field actions
as investigated by Koivisto with disformal quintessence~\cite{Koivisto2008}.

\subsection{Realistic FLRW cosmology}
In TeVeS, cold dark matter is absent. Therefore in order to get acceptable values for the physical Hubble constant today, we have to supplement
the absence of CDM with something else. Possibilities include the scalar field itself, massive neutrinos~\cite{SkordisEtAl2006,FerreiraSkordisZunkel2008}
and a cosmological constant. At the same time, one has to get the right angular diameter distance to  recombination~\cite{FerreiraSkordisZunkel2008}.
These two requirements can place severe constraints on the allowed free functions. 

\section{Linear cosmological perturbation  theory in TeVeS}
\label{sec_linear}

\subsection{Scalar modes}
The full linear cosmological perturbation theory in TeVes has been worked out by Skordis~\cite{Skordis2006} as well as for variants of TeVeS with
more general vector field actions~\cite{Skordis2008a}. The scalar modes of the linearly perturbed universally coupled metric are given in the conformal Newtonian gauge
as
\begin{equation}
ds^2 = -a^2(1+2\Psi) d\tau^2 + a^2 (1-2\Phi) \delta_{ij} dx^i dx^j
\end{equation}
where for simplicity we have assumed that the spatial curvature is zero and where $\tau$ is the conformal time defined as $dt = a d\tau$.
 See ~\cite{Skordis2006,Skordis2008a} for the curved cases, as well as
the cases of vector and tensor perturbations.
The scalar field is perturbed as $\phi = \bar{\phi} + \varphi$ where $\bar{\phi}$ is the FLRW background scalar field and $\varphi$ is the perturbation.
The vector field  is perturbed as $A_\mu = a e^{-\bar{\phi}}( \Psi -\varphi , \grad_i \alpha)$ 
such that the unit-timelike constraint is satisfied, which removes
the $A_0$ component as an independent dynamical degree of freedom.  Thus there are two additional dynamical degrees of freedom to GR, that is the scalar field
perturbation $\varphi$ and the vector field scalar mode $\alpha$.

The field equations for the scalar modes can be found in the conformal Newtonian gauge in~\cite{SkordisEtAl2006}, in general gauges in including the synchronous
gauge in~\cite{Skordis2006} and for more general TeVeS actions in~\cite{Skordis2008a}.

\subsection{Initial conditions for the scalar modes}
Setting up initial conditions for perturbations in cosmology has traditionally been classified in terms of adiabatic and isocurvature modes.
In the $\Lambda$CDM model five regular modes have been identified~\cite{BucherMoodleyTurok1999}, namely the adiabatic growing mode, the Baryon Isocurvature Density
mode, the CDM Isocurvature Density mode, the Neutrino Isocurvature Density mode and the Neutrino Isocurvature Velocity mode.

Generating initial conditions has always been  one of the most important issue in cosmology. One theory which generates initial conditions is inflation.
Typically single-field inflationary models predict an adiabatic spectrum of fluctuations, however, more general multi-field inflationary models or models
with curvatons predict a mixture of sometimes uncorrelated, and sometimes correlated,  adiabatic and isocurvature modes. Although generating the different
modes is important, the issue of their observability can be dealt with  separately, which in turn can place constraints on the theory which generates them.
Various multi-parameter studies of initial conditions and their observational impact on the CMB radiation and the LSS
have limited the total contribution from isocurvature modes to less than $30\%$ when all cases of arbitrarily correlated modes are allowed~\cite{BucherEtAl2004,MoodleyEtAl2004,DunkleyEtAl2005b} 
and to a few
percent in the case when an uncorrelated  single 
mode mixed with the adiabatic mode is allowed~\cite{CrottyEtAl2003,Kurki-SuonioMuhonenValiviita2004,BucherEtAl2004,MoodleyEtAl2004,BeltranEtAl2005,DunkleyEtAl2005b,BeanDunkleyPierpaoli2006,Trotta2006,KawasakiSekiguchi2007}. 
Thus in a $\Lambda$CDM cosmology, the dominant contribution must be adiabatic to a large extend.

The exact adiabatic growing mode in TeVeS and generalized variants has been found by Skordis in~\cite{Skordis2008a}, but only for the case of the 
generalized Bekenstein function. In general, the correct setup of initial conditions would depend on the free function. If, however, the free function
is such that the scalar field contribution to the background expansion, during the radiation era is very small, then the adiabatic mode for other free
functions would only marginally differ from the ones found in~\cite{Skordis2008a}. In particular, the only effect would be a difference in the initial
conditions of $\varphi$ and therefore it is unlikely that this would make any observational difference.

The only study of observational signatures of TeVeS theory in the CMB radiation and the LSS,
namely that of Skordis, Mota, Ferreira and B{\oe}hm~\cite{SkordisEtAl2006} have used initial conditions such  both $\varphi$ and $\alpha$ as well as their
derivatives are zero initially. While this is not a pure adiabatic initial condition, it turns out that it is close enough to the adiabatic initial conditions
found in~\cite{Skordis2008a} so that no observable difference can be seen from any isocurvature contamination.

Studies of isocurvature modes in TeVeS have not been conducted. In the light of the problems that TeVeS has with observations of the CMB
radiation~\cite{SkordisEtAl2006} it may be important to investigate what the observational effects are that the isocurvature modes would have.
For example correlated mixtures of adiabatic and isocurvature modes could lower the Integrated Sachs-Wolfe effect or raise the 3rd peak  both of which
pose significant problems to TeVeS. Preliminary studies by Mota, Ferreira and Skordis have shown that setting the vector field perturbations
large initially can have significant impact on both of these features~\cite{MotaFerreiraSkordis_edinburg}.

In addition to the four regular isocurvature modes that exist in GR as mentioned above, there could in principle exist four 
other isocurvature modes : two associated with the scalar field and two associated with the vector field. Preliminary studies by Skordis have shown that
none of the scalar field isocurvature modes are  regular~\cite{Skordis_preparation_2009a}. 
Conversely, under certain conditions of the vector field parameters one of the vector field isocurvature modes can be regular while the other one is never regular.
Thus, it may be possible to have one regular isocurvature mode in the TeVeS sector. The observational consequences of this mode are unknown as well as
its generation method from theories such as inflation in the presence of TeVeS fields.

\subsection{Linear equations for vector and tensor modes}
The full  linear cosmological perturbation formalism for vector and tensor modes has been worked out in~\cite{Skordis2006}  as well as for variants of TeVeS with
more general vector field actions~\cite{Skordis2008a}.  No studies of observable spectra based on vector or tensor modes have been conducted.

\subsection{Large Scale Structure observations}
\begin{figure}
\epsfig{file=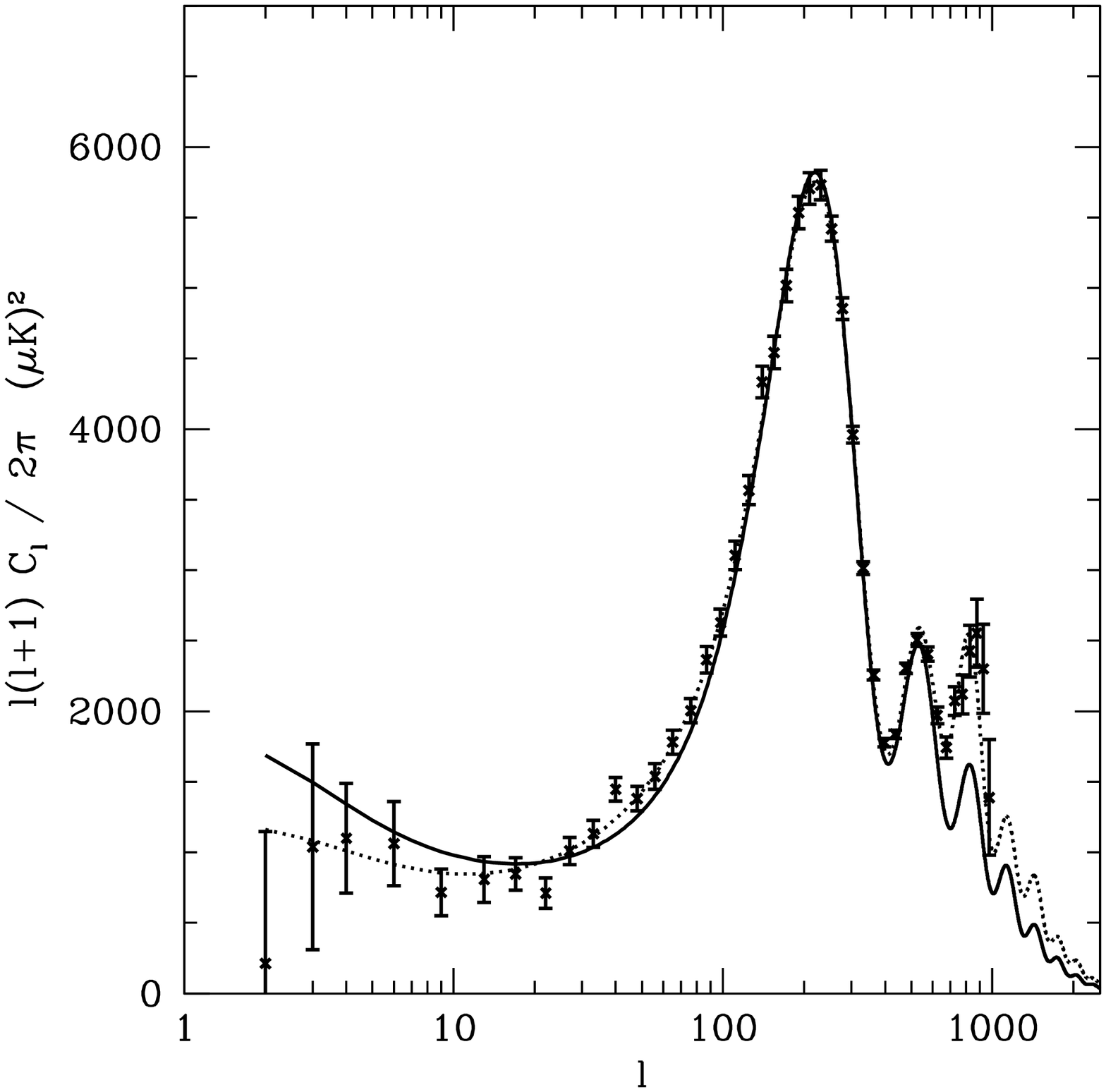,width=6.5cm}
\epsfig{file=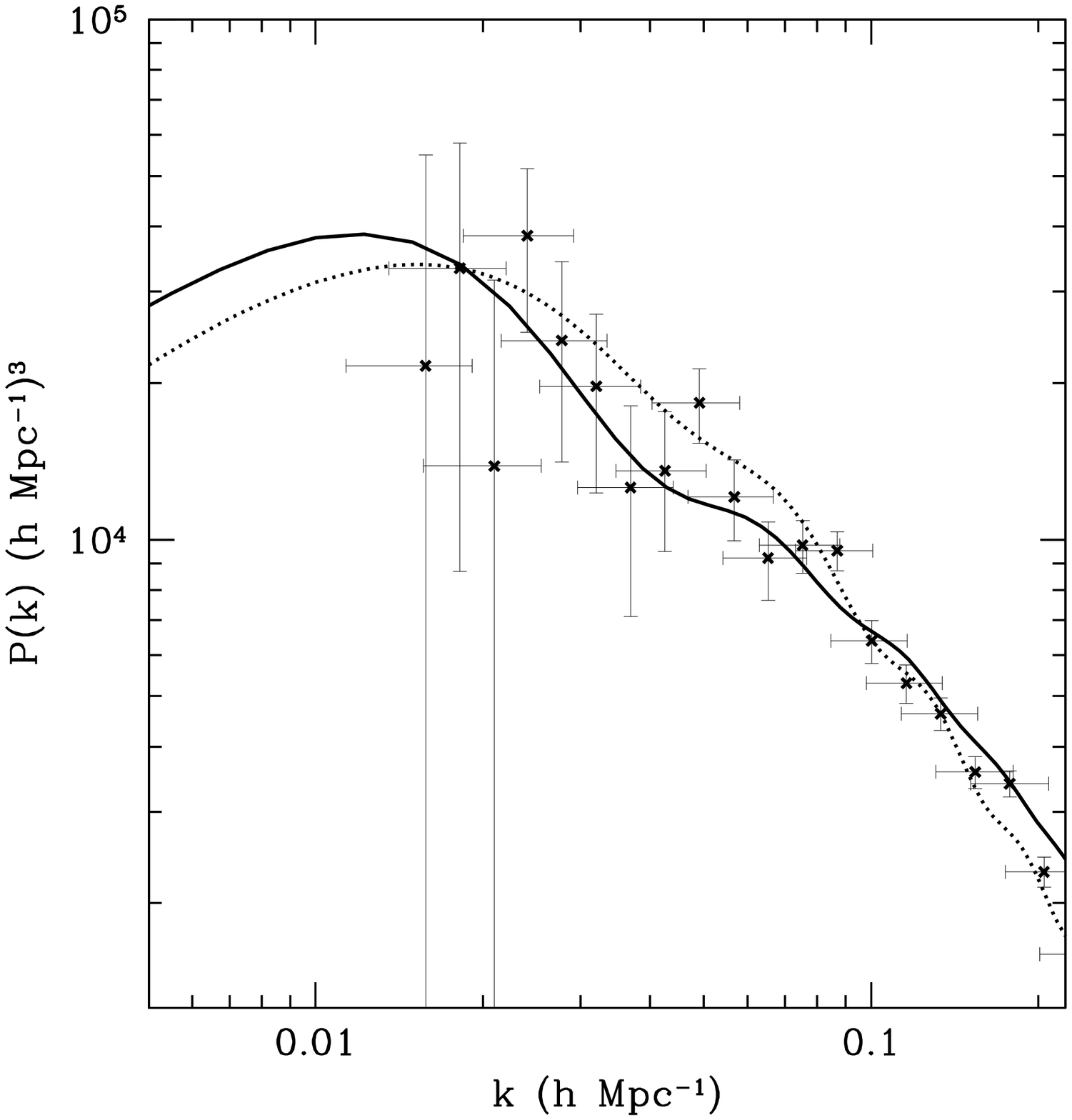,width=6.5cm}
\caption{
LEFT : The Cosmic Microwave Background angular power spectrum $l(l+1)C_l/(2\pi)$ for TeVeS (solid) and $\Lambda$CDM (dotted) with 
WMAP 5-year data~\cite{NoltaEtAl2008}.
\newline
RIGHT :The matter power spectrum $P(k)$ for TeVeS (solid) and $\Lambda$CDM (dotted) plotted with SDSS data.
\label{fig_Pk}
 }
\end{figure}
A traditional criticism of MOND-type theories was their lack of a dark matter component and therefore their presumed inability to form large scale structure
compatible with current observational data. This incorrect reasoning  was based on a General Relativistic universe filled only with baryons. In that case
it is well known that, since baryons are coupled to photons before recombination they do not have enough time to grow into structures on their own. 
Furthermore, their oscillatory behaviour at recombination is preserved and is visible as large oscillation in the observed galaxy power spectrum $P_{gg}(k)$. 
Finally, on scales smaller than the diffusion damping scale they are exponentially
damped due to the Silk-damping effect. CDM solves all of these problems because it does not couple to photons and therefore can start creating potential
wells early on, to which the baryons fall into. This is enough to generate the right amount of structure,  largely erase the oscillations and
overcome the Silk damping.

However, TeVeS is not General Relativity. It contains two additional fields, which change the structure of the equations significantly. 
The first study of TeVeS predictions for Large Scale Structure observations was conducted by  Skordis, Mota, Ferreira and B{\oe}hm~\cite{SkordisEtAl2006}.
They numerically solved the perturbed TeVeS equations for the case of the Bekenstein function and determined the effect on the matter power spectrum $P(k)$. 
They found that TeVeS can indeed form large scale structure compatible with observations depending
on the choice of TeVeS parameters in the free function. In fact the form of the matter power spectrum $P(k)$ in TeVeS looks  quite similar to that in
$\Lambda$CDM. Thus TeVeS can produce matter power spectra that cannot be distinguished from $\Lambda$CDM. One would have to turn to other observables
to distinguish the two models. The power spectra for TeVeS and $\Lambda$CDM are plotted on the right panel of Figure \ref{fig_Pk}.

Dodelson and Liguori~\cite{DodelsonLiguori2006} provided an analytical explanation of the growth of structure seen numerically by~\cite{SkordisEtAl2006}.
 They have found that the growth in TeVeS cannot  be  due to the scalar field. In fact the scalar field
perturbations have Bessel function solutions and are decaying in an oscillatory fashion. Instead, they found that the growth in TeVeS is due to the
vector field perturbation.

Let us see how the vector field leads to growth.  Using the tracker solutions in the matter era from Bourliot \etal~\cite{BourliotEtAl2006}
we find the behaviour of the background functions $a$,$ b$ and $\bar{\phi}$. These are used into the perturbed field equations, 
after setting the scalar field  perturbations to zero, and we find that in the matter era the vector field scalar mode $\alpha$ obeys the equation
\begin{equation}
\ddot{\alpha} +  \frac{b_1}{\tau} \dot{\alpha}
+ \frac{b_2}{\tau^2} \alpha 
=  S(\Psi,\dot{\Psi},\theta)
\label{eq_vec_growth}
\end{equation}
in the conformal Newtonian gauge, where 
\begin{eqnarray}
b_1&=&\frac{4(\mu_0\mu_a-1)}{\mu_0\mu_a+ 3}
\\
b_2 &=& \frac{2}{(\mu_0\mu_a + 3)^2}    \left[ \mu_0^2\mu_a^2 - \left(5+\frac{4}{K}\right)\mu_0\mu_a +  6  \right] .
\end{eqnarray}
and where $S$ is a source term which does not explicitly depend on $\alpha$.
If we take the simultaneous limit $\mu_0\rightarrow\infty$ and $K\rightarrow 0$
for which $\Omega_\phi \rightarrow 0$ meaning that the TeVeS contribution is absent, we get $b_1 \rightarrow 4$ and $b_2 \rightarrow 2$.
In this case the two homogeneous solutions to  (\ref{eq_vec_growth}) we $\tau^{-2}$ and $\tau^{-1}$ which are decaying. Dodelson and
Liguori show that the source term $S(\Psi,\dot{\Psi},\theta)$ is not sufficient to create a growing mode out of the general solution to
(\ref{eq_vec_growth}) and therefore in this General Relativistic limit, TeVeS does not provide enough growth for structure formation.

\begin{figure}
\epsfig{file=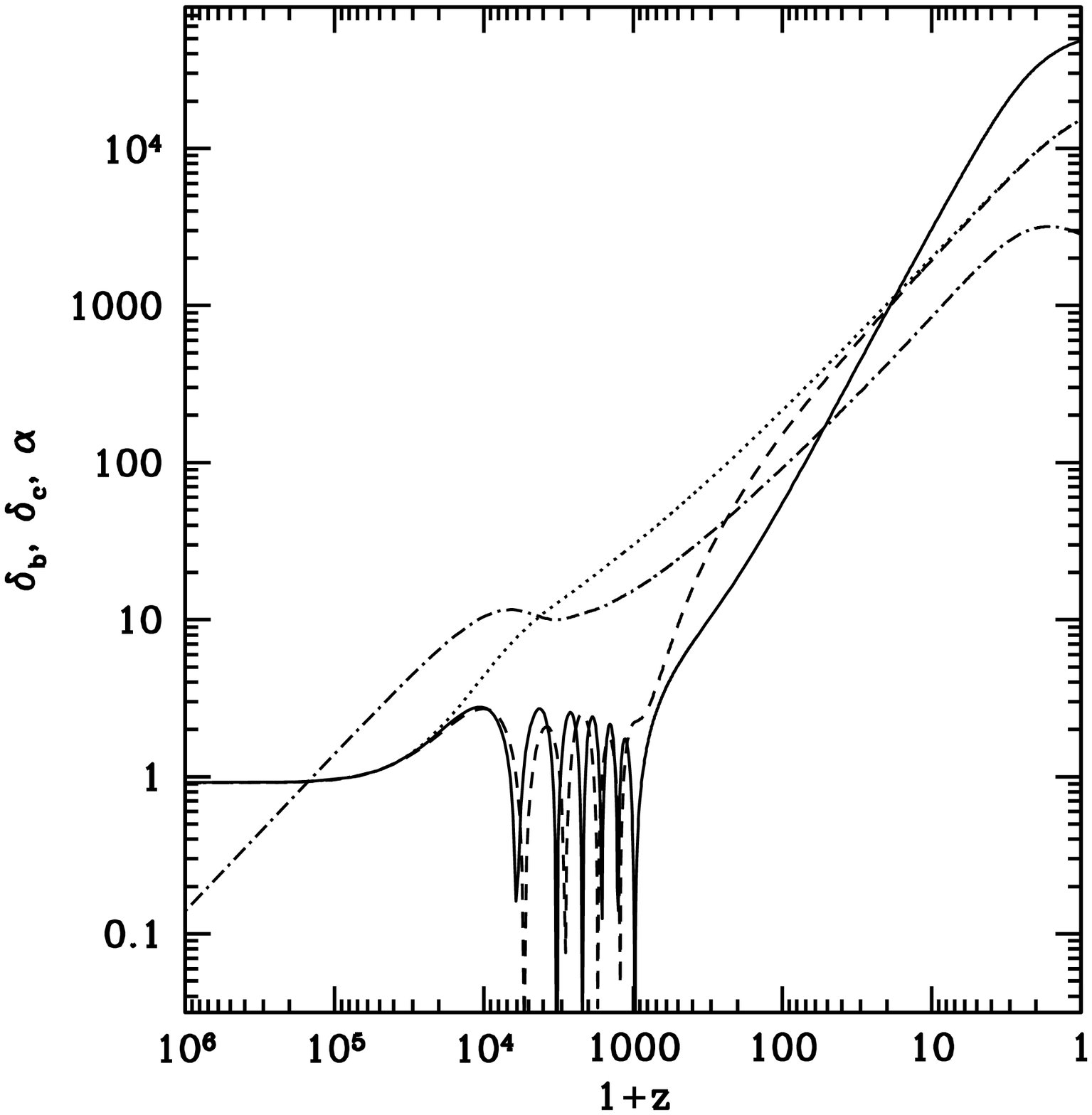,width=6.5cm}
\epsfig{file=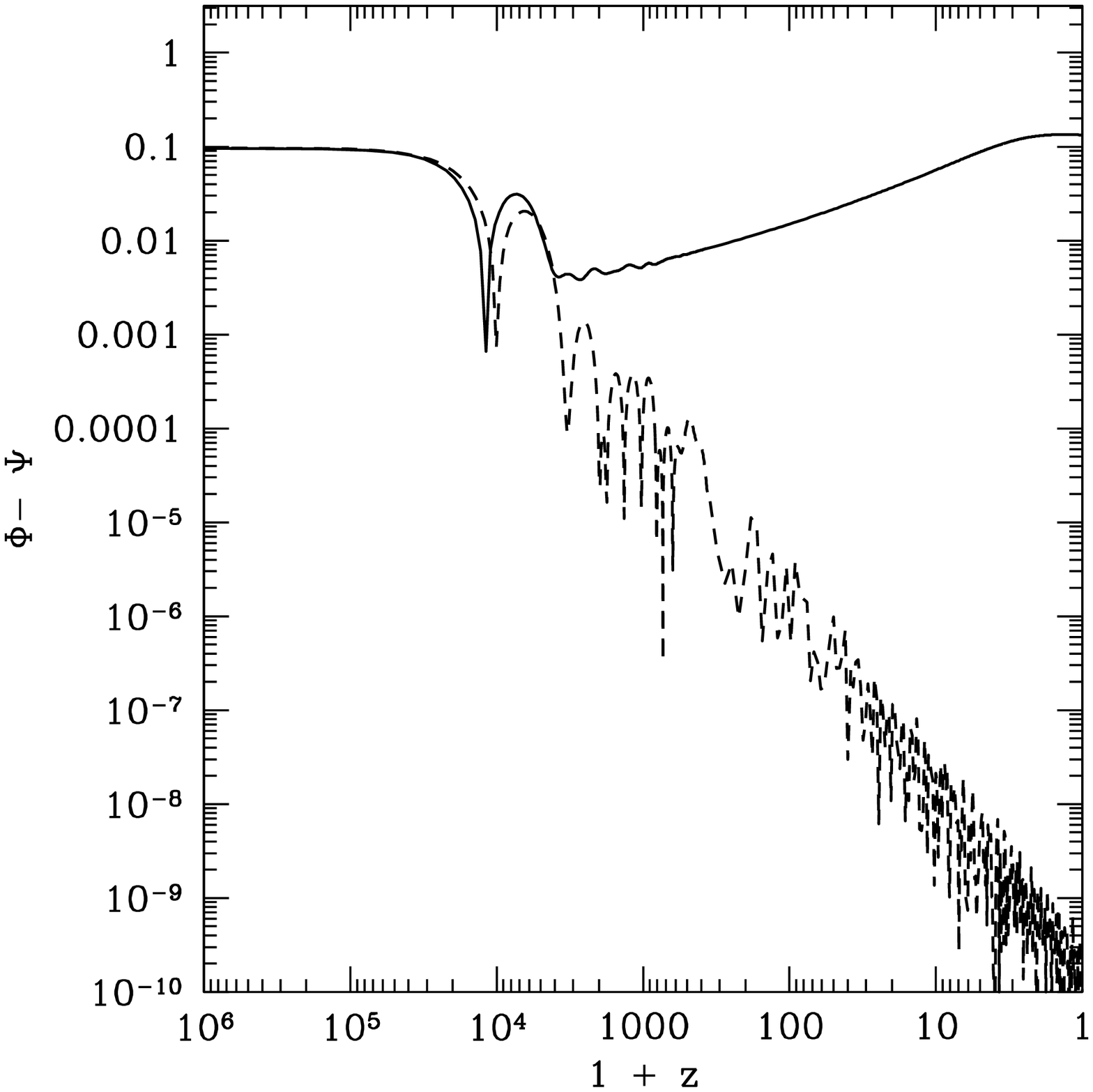,width=6.5cm}
\caption{
LEFT: The evolution of the baryon density fluctuation in TeVeS (solid) and $\Lambda$CDM (dashed) with redshift for
a wavenumber $k=10^{-3}Mpc^{-1}$. Notice how in both cases, baryons fluctuate before recombination and grow afterwards.
In the case of $\Lambda$CDM, the baryons eventually follow the CDM density fluctuation (dotted) which starts growing before
recombination. In the case of TeVeS, the baryons grow due to the potential wells formed by the growing scalar mode in
the vector field $\alpha$ (dash-dot).
\newline
RIGHT: The difference of the two gravitational potentials $\Phi - \Psi$ for a wavenumber $k=10^{-3}Mpc^{-1}$ plotted against redshift.
The solid curve is a TeVeS model which gives the correct matter power spectrum. 
 The dotted curve is a  $\Lambda$CDM model.
}
\label{fig_phi_psi}
\end{figure}

Now lets look at the general case. Dodelson and Liguori assume the ansatz 
that the homogeneous solutions to (\ref{eq_vec_growth})  are given as $\tau^n$ for some power
index $n$. Generalizing their result to the generalized Bekenstein function of Bourliot \etal~\cite{BourliotEtAl2006} we  get that
\begin{equation}
n = \frac{-3 + \frac{7}{\mu_0\mu_a} \pm  \sqrt{ 1+ \frac{1}{\mu_0^2\mu_a^2}   -\frac{2}{ \mu_0\mu_a}
 +\frac{32}{K\mu_0\mu_a }
}}{2 (1 + \frac{3}{\mu_0\mu_a})}
\end{equation}
Now since $\Omega_\phi \sim 10^{-3}$ we have that $\frac{1}{\mu_0\mu_a} \sim 3 \times 10^{-3}$ and we may ignore these terms to get
\begin{equation}
n \approx -\frac{3}{2} + \frac{1}{2}  \sqrt{ 1 +\frac{32}{K\mu_0\mu_a  } }
\end{equation}
where we have ignored the negative sign which would give a decaying mode.
Thus we can get $n>0$ provided 
\begin{equation}
K < \sim  0.01
\end{equation}
for fixed $\mu_0\mu_a$. Smaller values of $\mu_0\mu_a$ can also raise this threshold.
Thus if this condition is met, there exists a growing mode in $\alpha$. This in turn feeds back  into the 
perturbed Einstein equation and sources the gravitational potential $\Phi$. 
This is translated into a non-decaying mode in $\Phi$ which in turn drives structure formation.
This is graphically displayed in the left panel of Figure ~\ref{fig_phi_psi}.

It is a striking result that even if the contribution of the TeVeS fields to the background FLRW equations is negligible ($\sim 10^{-3}$ or less),
we can still get a growing mode which drives structure formation.
This explains analytically the numerical results of  Skordis, Mota, Ferreira and B{\oe}hm~\cite{SkordisEtAl2006}.

\subsection{Cosmic Microwave Background observations}
A General Relativistic universe dominated by baryons cannot fit the most up to date observations of the CMB anisotropies~\cite{NoltaEtAl2008}.
This is true even if a cosmological constant and/or three massive neutrinos are incorporated into the matter budget so that 
the first peak of the CMB  angular power spectrum is at the right position. Although a model with baryons, massive neutrinos
and cosmological constant can give the correct first and second peaks, it gives a third peak which is lower than the second. On the contrary,
the measured third peak is almost as high as the second. This is possible if CDM is present.

Once again, TeVeS is not General Relativity and it is premature to claim (as in~\cite{SlosarMelchiorriSilk2005,SpergelEtAl2006}) 
that only a theory with CDM can fit CMB observations.  In fact as was shown by Ba{\~n}ados, Ferreira and Skordis~\cite{BanadosFerreiraSkordis2008}, the
Eddington-Born-Infeld (EBI) theory~\cite{Banados2008,BanadosEtAl2008} can produce CMB spectra that look like $\Lambda$CDM. Although the EBI theory has no relation to TeVeS and has features which 
are rather different than TeVeS, it is not inconceivable that TeVeS theory or some variant could also give similar results.
  Skordis, Mota, Ferreira and B{\oe}hm~\cite{SkordisEtAl2006}
numerically solved the linear Boltzmann equation in the case of TeVeS and calculated the CMB angular power
spectrum for TeVeS. By using initial conditions close to adiabatic the spectrum thus found provides very poor fit as compared to
the $\Lambda$CDM model (see the left panel of Figure \ref{fig_Pk}). The  CMB seems to put TeVeS into trouble, at least for the Bekenstein free function.
If this was a complete study then TeVeS would already be ruled out by the CMB data.

It may be that different variants of TeVeS, e.g. with different vector field actions, or different scalar free functions
could give better fits, but at the moment this is still an open problem. A different possibility is the use of correlated isocurvature modes,
in particular modes from the TeVeS sector. This will undoubtedly give different spectra but once again the question of whether it will lead to
acceptable spectra is open.

Needless to say, if none of the aforementioned possibilities works it would quite possibly be the end of TeVeS as a theory capable of explaining the 
mass and energy discrepancies on galactic, cluster and cosmological scales.

We should note however, that the presence of a fourth sterile neutrino with mass $\sim 11 eV$ will give very good fits to the  CMB 
angular power spectrum, as found by Angus~\cite{Angus2008b}. This is true also in General Relativity and so this result has nothing to do with
TeVeS features. However, in General Relativity this neutrino would severely suppress Large Scale Structure, and only in TeVeS there is the possibility
to counteract it.

\subsection{Sourcing a difference in the two gravitational potentials : $\Phi - \Psi$}
The result of Dodelson and Liguori~\cite{DodelsonLiguori2006} has a further direct consequence. The perturbed TeVeS equations which
relate the difference of the two gravitational potentials $\Phi -\Psi$ to the shear of matter, have additional contributions coming from the 
perturbed vector field $\alpha$. This is not due to the existence of the vector field  per se but comes from the disformal transformation to which the
vector field plays a fundamental role. Indeed in a single metric theory where the vector field action is Maxwellian (just like TeVeS), there is no
contribution of the vector field to $\Phi - \Psi$.

Since the vector field is required to grow in order to drive structure formation, it will inevitably  lead to a growing $\Phi - \Psi$. This is 
precisely what we see numerically in the right panel of Figure~\ref{fig_phi_psi}.

If the difference $\Phi - \Psi$, named the gravitational slip,
 can be measured observationally, it can provide a substantial test of TeVeS that can distinguish TeVeS from $\Lambda$CDM.
The link between $\Phi-\Psi$ and theories of gravity has been noted before by Lue, Starkman and Scoccimarro~\cite{LueScoccimarroStarkman2004} and
by Bertschinger~\cite{Bertschinger2006}. Since Dodelson and Liguori uncovered its importance for theories like TeVeS,
 many authors~\cite{CaldwellCoorayMelchiorri2007,DoreEtAl2007,JainZhang2007,SchmidtLiguoriDodelson2007,ZhangEtAl2007,DanielEtAl2008} have
investigated various observational techniques to probe the gravitational slip. Theoretical consideration on general tests of modifications to gravity and
their effect on observations have been investigated in~\cite{HuSawicki2007a,Hu2008,Skordis2008b}. 

\subsection{Inflation and TeVeS}
In TeVeS, the disformal transformation plays a fundamental role. It may be that this can lead to interesting features in the
initial power spectrum and/or the production of isocurvature modes in the TeVeS sector, in an inflationary universe. 
Disformal inflation has been investigate before by Kaloper~\cite{Kaloper2003} but inflation in the context of TeVeS remains an unexplored direction.
If isocurvature perturbations are found to significantly contribute to the CMB anisotropies, it would be
of particular importance to check whether they can be produced in an inflationary era in TeVeS.

\section{Non-cosmological studies of TeVeS theory}
\label{sec_noncosmo}
It is not the main focus of this review to present non-cosmological studies of TeVeS theory, however,
in this section I give a brief outline of the work done along a variety of directions. This includes
spherically symmetric systems, such as Black Holes and neutron stars, galactic rotation curves,
gravitational lensing, galaxy clusters and galaxy groups, post-Newtonian parameters and theoretical issues such as
stability and singularities.

\subsection{Spherically symmetric systems}

\subsubsection{Black holes}
Static, spherically symmetric vacuum systems were first considered by Bekenstein in the original TeVeS paper~\cite{Bekenstein2004a},
in order to calculate some of the PPN parameters for TeVeS. 
The first spherically symmetric static vacuum solutions in TeVeS, however, were thoroughly studied by Giannios~\cite{Giannios2005}.
Under the assumption of strong field limit $\mu = \mu_0$,
he found a family of vacuum solutions described in isotropic coordinates by the physical metric
\begin{equation}
ds^2 = - \left(\frac{r-r_c}{r+r_c}\right)^n dt^2 + \frac{(r^2 - r_c^2)^2}{r^4}   \left(\frac{r-r_c}{r+r_c}\right)^{-n}  \left(dr^2 + r^2 d\Omega^2 \right)
\end{equation}
where $r_c$ is an integration constant of dimensions of length and the power $n$ is given by
\begin{equation}
 n = 2\sqrt{\frac{1 - \frac{1}{2\mu_0} \frac{GM_s}{r_c}}{1-\frac{K}{2}}} + \frac{2}{\mu_0} \frac{GM_s}{r_c}.
\label{eq_n_BH}
\end{equation}
The mass $M_s$, called the scalar mass, is determined in the case where the above solution is an exterior solution to a star as
\begin{equation}
M_s =  \int d\Omega \int_0^R dr \sqrt{-g}(\rho + 3 P)
\label{eq_scalar_mass}
\end{equation}
In the above solution, it was assumed that the vector field is aligned with the time direction and is thus fixed by the unit-norm
constraint. The scalar field was found to have the profile
\begin{equation}
\phi = \phi_c +   \frac{GM_s}{\mu_0r_c} \ln\frac{r-r_c}{r+r_c}
\end{equation}

Although the above solution describes an exterior solution, it does not necessarily describe a Black Hole. This is to be expected since
there is no  Jebsen-Birkhoff~\cite{Jebsen1921,Birkhoff1923} theorem in TeVeS due to the presence of additional fields to the metric. Thus 
vacuum spherically symmetric solutions are not unique. For the above solution to describe a Black Hole, the candidate event horizon
at $r=r_c$ must have bounded surface area (which implies $n\le 2$) and not lead to an essential curvature singularity (which implies
$n=2$ or $n>4$)~\cite{SagiBekenstein2008}. These two conditions taken together imply that Black Holes in TeVeS require $n=2$, in which
case the metric above becomes exactly the  Schwartzschild metric. This also allows us to determine $M_s$ in terms of $r_c$, $\mu_0$ and $K$ from (\ref{eq_n_BH}).

There is one caveat that results from this solution, that is for small enough $r$ but still greater than $r_c$, the scalar field
can acquire negative values, i.e. since $\frac{r - r_c}{r+r_c} < 1$ for all $r>r_c$, there is always some radius $r_1(\phi_c)$ such that
$\phi<0$ for $r<r_1$. As shown by Bekenstein~\cite{Bekenstein2004a} this is sufficient to lead to superluminal propagation
of perturbations when viewed with the physical metric. Thus very close to the Black Hole it would be possible to create
closed signal curves.

It would thus seem that Black Holes in TeVeS are unphysical. However, in the light of the absence of a Jebsen-Birkhoff theorem,
there is also the possibility that there exists a different Black Hole solution which does not allow for the scalar field to
become negative and is thus physical.

Indeed this other solution branch has been found by Sagi and Bekenstein as a by-product while studying charged Black Hole
solutions~\cite{SagiBekenstein2008}. Sagi and Bekenstein made an educated guess that a charged Black Hole in TeVeS has
 the Reissner-Nordstr\"{o}m physical metric, 
\begin{eqnarray}
ds^2 &=& -\frac{ (r^2 -r_h^2)^2 }{(r^2  + G_N Mr + r_h^2)^2} dt^2 
\nonumber 
\\
&& + \frac{(r^2 + G_N Mr + r_h^2)^2}{r^4} \left(dr^2  + r^2 d\Omega^2\right)
\end{eqnarray}
where $M$ is the  physical mass   of the black hole and not the mass-parameter used in the original work~\cite{SagiBekenstein2008}, 
$r_h$ is a scale given by $4r_h^2 = G_N(G_NM^2 - Q^2)$ 
and $Q$ is the physical Black Hole charge~\footnote{It might not be directly apparent that $M$ and $Q$ are the
physical mass and charges of the Black hole. It is not hard, however, to show that they are as was done by Sagi and Bekenstein~\cite{SagiBekenstein2008}.}. 
Remember that the observed Newton's constant $G_N$ is related to $G$ through (\ref{eq_G_GN}).
Note also that I'm using rescaled units  from~\cite{SagiBekenstein2008}
such that no $e^{\phi_c}$ appear in the physical metric. This is done by the 
usual transformation $t\rightarrow e^{-\phi_c}t$ and $r\rightarrow e^{\phi_c} r$ 
(thus one has to also transform all relevant dimensionful variables e.g. $r_h \rightarrow  e^{\phi_c}  r_h$).

The above metric uniquely determines the solution for the electic field  as usual.
This enabled them to find solutions for the scalar field and
for the vector field under the assumption that it points to the time direction. It turns out that the scalar field equation permits two branches of solutions.
One of them, corresponds to the Giannios solution in the $Q\rightarrow 0$ limit which allows for superluminal propagation, a conclusion also unchanged in the charged case.
The other branch, however, does not have the problem of superluminality. In that case the scalar field is given by
\begin{eqnarray}
\phi &=& \phi_c  +   \delta_+ \ln(1+\frac{r_h}{r})
\nonumber 
\\
&&
 + \delta_- \ln (1 - \frac{r_h}{r}) 
- \frac{1}{2}(\delta_+ + \delta_-) \ln(1 + \frac{G_NM}{r} + \frac{r_h^2}{r^2}) 
\label{eq_phi_CBH}
\end{eqnarray}
where the two constants $\delta_+$ and $\delta_-$ are given by
\begin{equation}
\delta_{\pm} = \frac{2-K \pm \sqrt{ 2(2-K) + \mu_0 K } }{2-K + \mu_0 }
\end{equation}
Clearly $\delta_+ \ge 0 $ and $\delta_- \le 0$, therefore the term $\delta_- \ln (1 - \frac{r_h}{r})$ is always positive which means that for $\phi_c>0$ (which would
also make sense cosmologically) we have $\phi>0$ and superluminal propagation is always avoided close to the black hole. Away from the black hole the situation can change.
Expanding (\ref{eq_phi_CBH}) to $O(\frac{1}{r})$ as $r\rightarrow \infty$ we get that $\phi$ approaches the asymptote $\phi_c$ from above iff
\begin{eqnarray}
\frac{ G_NM}{2r_h} \le  \frac{\delta_+ - \delta_-}{ \delta_+ + \delta_-} = \frac{ 2\sqrt{ 2(2-K) + \mu_0 K } }{ 2-K}
\end{eqnarray}
which is the condition found in~\cite{SagiBekenstein2008} in a different manner. Thus if the condition above is satisfied then superluminal propagation is avoided
everywhere for $\phi_c >0$. On the other hand violation of the above inequality can lead to superluminal propagation unless a sufficiently large value for $\phi_c$
is used. Since we expect $\phi_c$ to be small, this limits the possibilities. In particular for extremal Black Holes, for which $r_h=0$ we can never satisfy the 
above condition.

Taking the limit $Q\rightarrow 0$ (for which $G_NM=2r_h=2r_c$) we get the Giannios solution but with a negative $M_s$. If the above solution 
was the exterior solution to a star then it would have been discarded. Since however it is a black hole, $M_s$ is no longer determined by (\ref{eq_scalar_mass}) but
only the ratio $\delta_- = \frac{2}{\mu_0}\frac{G}{G_N}\frac{M_s}{M} \le0$ which is determined in terms of the TeVeS parameters is important.
 As a consequence, it is not clear how such a black hole (even when $Q=0$)
could arise from gravitational collapse, and this could be the subject of a future investigation.

Sagi and Bekenstein conclude their paper with the thermodynamics of these charged Black Holes and find that the usual Black Hole theormodynamics
for the physical metric are recovered in TeVeS. It has been shown by Dubovsky and Sibiryakov~\cite{DubovskySibiryakov2006} that theories with
Lorentz symmetry breaking via a time-dependent scalar field, in which there is more than one maximal propagation speed, could lead to
perpetual motion machines. It was also found by Eling \etal~\cite{ElingEtAl2007} that classical violation of the 2nd law of thermodynamics is
also possible in Einstein-{\AE}ther theory, also relying on two maximal propagation speeds. Given that TeVeS has both of these characteristics, i.e.
local Lorentz symmetry violation and different maximal propagation speeds for electromagnetic and gravitational waves, does it give rise to
these problems?  Sagi and Bekenstein answer this question to the negative.

\subsubsection{Violation of the Jebsen-Birkhoff theorem in TeVeS.}
It is obvious from the different families of vacuum solutions found in TeVeS theory that the
Jebsen-Birkhoff theorem is violated. In a way this was to be expected due to the presence of additional gravitational fields to the metric.
 As shown by Dai, Matsuo and Starkman~\cite{DaiMatsuoStarkman2008a}, 
the consequences of the violation of the Jebsen-Birkhoff theorem can be dramatic and can complicate the observability of such theories.
This means there are no truly isolated objects in relativistic MOND theories as even a small perturbation can lead to large violations of the 
theorem.
Further studies should be conducted, however, to establish the extend of which this can influence observations in TeVeS theory and related variants.

\subsubsection{Neutron stars}
Jin and Li considered the Tolman-Oppenheimer-Volkoff~\cite{Tolman1939,OppenheimerVolkoff1939} equation in the context of TeVeS under the assumption of
strong-field limit $\mu = \mu_0$ and constant energy density inside the star~\cite{JinLi2006} 
and solved the equations perturbatively in terms of the distance from the centre of the star.

Lasky, Sotani and Giannios made a much more thorough study, by considering neutron stars~\cite{LaskySotaniGiannios2008},
again under the assumption $\mu=\mu_0$ (strong-field limit). They numerically solved the Tolman-Oppenheimer-Volkoff
equations for hydrostatic equilibrium inside a spherically symmetric star with the realistic polytropic equation of state
of Pandharipande~\cite{Pandharipande1971} ( equation of state "A" of Arnett and Bowers~\cite{ArnettBowers1977}).  
They further assumed that both the vector field  and fluid velocity are aligned with the 
time direction and are thus proportional ($u_a = e^\phi A_a$) in which case the vector field equations are identically satisfied.
 Not surprisingly they find that sensible solutions are possible only  if $K<2$, which is indeed what is implied by positivity of energy 
and positivity of the effective Newton's constant (\ref{eq_G_GN}). 

The interior neutron star solution is matched to the exterior  Schwartzschild-TeVeS solution
found by Giannios~\cite{Giannios2005} which is then used to define an Arnowitt-Deser-Misner (ADM) mass for the star. This last assumption, however,
is a point that should warrant further investigation in the future. As discussed above, the  Schwartzschild-TeVeS solution is valid only 
in the $\mu=\mu_0$ limit
and thus it is correct only close to a black hole, or the surface of the star. Far away from the star, as the potential gradient drops, the solution would eventually
move away from $\mu=\mu_0$ and, in a way that would depend on the TeVeS function $V(\mu)$, settle towards the MOND regime. The exception to this would be
if the star is situated in an external field such as a galaxy, which itself is in the Newtonian regime.
 The MOND regime
cannot lead to an asymptotically flat solution, at least not for the matter-frame metric, and thus the meaning of the ADM mass is questionable. Nevertheless,
it is still a parameter that can be used to study these solutions. 

By varying the two TeVeS parameters, $K$ and $\mu_0$,  they determined how the ADM mass parameter depends on the radius of the star.
Both the ADM mass and the radius of the star can be significantly different from the ones in General Relativity.
Taking an indicating upper bound for the masses of neutron stars as $M\sim1.5 M_\odot$, they find that $K$ is conservatively constrained to be at least less than unity,
which is greater than typical values required by cosmology.
They also consider the variation of the scalar field inside the star and find that depending on its initial cosmological value $\phi_c$, it can become negative in
the interior. As found by Bekenstein~\cite{Bekenstein2004a} such a situation leads to superluminal propagation of perturbations which could
be used to create closed signal curves~\cite{BonvinCapriniDurrer2007}. This can be avoided provided $\phi_c > 10^{-3}$.

In the final section of the paper, the authors discuss possible observational signatures. They note that the redshift of atomic spectral
lines emanating from the surface of the star depends on the TeVeS parameters. As discussed by DeDeo and Psaltis~\cite{DeDeoPsaltis2003},
measurements of these lines could in turn place constraints on gravitational theories (and therefore TeVeS theories and their variants )
using X-ray observatories such as Chandra~\cite{SanwalEtAl2002} and XMM-Newton~\cite{CottamPaerelsMendez2002}.
A second test that Lasky, Sotani and Giannios propose is to probe TeVeS with gravitational wave astronomy, by determining the compactness of
neutron stars through the use of w-mode oscillations~~\cite{KokkotasSchutz1992}, as these are almost independent of the equation of state of the neutron
star~\cite{KokkotasSchmidt1999}.

 As they also point out, studies of perturbations would also lead to predictions for quasi-periodic oscillations of neutron stars, which has also
been proposed by DeDeo and Psaltis~\cite{DeDeoPsaltis2004} as a possible test of gravitational theories.  The quasi-periodic oscillations in 
the giant flares  produced by magnetars  (magnetized neutron stars)~\cite{IsraelEtAl2005,StrohmayerWatts2005,StrohmayerWatts2006}
seem to be in good aggreement with General Relativity~\cite{SotaniKokkotasStergioulas2007}, and would therefore provide strong tests for TeVeS and similar theories.
 Further studies of neutron stars could also involve adding a radial part to the vector field or choosing a different equation of state.

In a different study, Sotani~\cite{Sotani2009}  has analyzed the oscillation frequencies of neutron stars in TeVeS  in the Cowling approximation. 
That is, he assumed that the fluctuations of the three gravitational fields $\metE_{ab}$, $\phi$ and $A_a$ 
are frozen and considered only the fluctuations of the fluid. He then considered two different equations of state and determined the dependence of
the frequency of the f-mode oscillations (fundamental modes) as a function of the stellar averaged density. It turns out that the result depends predominantly on
the TeVeS constant $K$ and that the dependence on the equation of state is weak. When $K$ increases, the frequency of oscillation increases with larger average stellar masses.
Observing such oscillations can then put constraints on the constant $K$.

\subsubsection{Stability of spherically symmetric perturbations.}
Related to the spherically symmetric solutions discussed above is the issue of their stability to spherically symmetric perturbations.
Seifert and Wald have formulated a  variational method, based on earlier work of Wald~\cite{Wald1990a}, for determining the stability of spherically symmetric perturbations around
a background spherically solution for any metric-based theory of gravity~\cite{SeifertWald2006}. Seifert applied~\cite{Seifert2007} this method to
TeVeS, amongst other theories. He found that in the case of TeVeS the variational method is inapplicable because it does not lead to a
well defined inner product on the space of perturbational fields. Using a WKB approximation, however, he was able to show that the  
branch of the  Schwartzschild-TeVeS solution found by Giannios~\cite{Giannios2005}, where the vector field is aligned with the time direction, is unstable. 

The question of  stability of other solutions, such as the
charged black hole and the 2nd branch of the Schwartzschild-TeVeS solution found by Sagi and Bekenstein~\cite{SagiBekenstein2008}, or the neutron star solutions
found by Lasky, Sotani and Giannios~\cite{LaskySotaniGiannios2008} 
is still open. It is also an open question whether other TeVeS variants
 could provide for stable solutions. This is certainly a possibility, as Seifert has demonstrated that
parameter spaces exist in  general Einstein-{\AE}ther theories for which the spherically symmetric solutions are stable. Moreover as shown by
Zlosnik~\cite{Zlosnik_Thesis} vacuum solutions in a pure AQUAL theory are stable. 
These issues are currently under investigation~\cite{SkordisZlosnik_preparation2009}. 

Before concluding this section, a brief comment is in order. Whether some solutions are unstable does not
necessarily mean that there are no stable solutions but rather that the unstable solution cannot be realized in nature. The issue of the existence
of stable solutions is  a different matter all together.

\subsection{General stability and gravitational collapse}
Contaldi, Wiseman and Withers~\cite{ContaldiWisemanWithers2008} investigated the gravitational collapse of boson stars in TeVeS.
They found that quite generically, the present TeVeS action leads to caustic formation and naked singularities before a Black Hole horizon can form.
They trace the source of the caustic formation to the vector field and the particular form of its action.
This is a serious theoretical obstacle to TeVeS. However, as they point out, slight modifications of the vector field action can introduce terms which
prevent caustic formation in the absence of the scalar field. 

A full investigation of caustic formation with  general vector field action in TeVeS 
and with the full inclusion of the scalar field is still lacking. Their results show that it is an important issue to be addressed.

\subsection{Parameterized Post-Newtonian parameters and tests of TeVeS in the solar system}
The first calculation of the PPN parameter $\beta^{(PPN)}$ and $\gamma^{(PPN)}$ was done by Bekenstein~\cite{Bekenstein2004a},
where he showed that $\gamma^{(PPN)} = 1$ and  $\beta^{(PPN)} = 1$~\footnote{The  $\beta^{(PPN)}$ parameter in~\cite{Bekenstein2004a} is 
incorrect. The correct $\beta^{(PPN)}$ can by found in the erratum~\cite{Bekenstein2005Err} with thanks to Giannios (see ~\cite{Giannios2005}).}.
Thus TeVeS gives identical predictions to General Relativity for these two parameters.

The (almost) full set of PPN parameters were calculated by Tamaki~\cite{Tamaki2008} using Will's procedure~\cite{Will1981,FosterJacobson2005}. 
He verified Bekenstein's and Giannios's calculation of $\beta^{(PPN)}$ and $\gamma^{(PPN)}$,
and further found that $\xi^{(PPN)}=\zeta^{(PPN)}_i = \alpha^{(PPN)}_3=0$. 
His method, however, did not allow him to calculate the very important prefered frame parameters $\alpha^{(PPN)}_1$ and $\alpha^{(PPN)}_2$.
The complete set of PPN parameters (including  $\alpha^{(PPN)}_1$ and $\alpha^{(PPN)}_2$) were finally calculated by Sagi~\cite{Sagi_preparation_2009}.

The PPN parameters determine corrections to Newtonian gravity towards the strong-field regime and
cannot determine deviations towards the MOND regime. These deviations could be important and detectable in the solar system.
For example, one place where such deviations could play a role is the Pioneer anomaly~\cite{TuryshevNietoAnderson2005}. Such possibilites
are discussed by Sanders~\cite{Sanders2006}.
Calculating MOND corrections in TeVeS theory (and indeed in any relativistic MOND theory) is still an open problem, although
a first attempt was initiated by Bonvin \etal~\cite{BonvinEtAl2008} for the case of the generalized Einstein-{\AE}ther theory.
Further solar system tests  of relativistic MOND theories could also be performed using different techniques~\cite{BekensteinMagueijo2006}.

\subsection{Gravitational lensing}
Gravitational lensing has become a crucial observational tool in modern Cosmology. It is often cited as giving the most
compelling evidence for the existence of Dark Matter. Objects like the Bullet cluster rely on both strong and weak gravitational lensing
to map the gravitational potential wells to show that they are displaced from the majority of the visible baryon mass. It is thus
of particular importance to determine the TeVeS predictions on gravitational lensing.

A first calculation of the deflection of light in TeVeS has been done by Bekenstein~\cite{Bekenstein2004a}. He showed that in the weak
acceleration regime TeVeS provides the right deflection of light as if dark matter was present. The vector field plays a crucial role in the 
derivation through the disformal transformation. Indeed that was precisely the reason for introducing the vector field into the theory in the 
first place by Sanders~\cite{Sanders1997}.
Various authors have tested TeVeS with gravitational lenses and a
short introduction of gravitational lensing in TeVeS can be found in~\cite{Zhao2006b}.

Theoretical aspects of gravitational lensing in TeVeS have been discussed by Chiu, Ko and Tian~\cite{ChiuKoTian2005} 
who looked at pointlike masses. They point out that the difference in amplifications for two images coming from a distant source
 is no longer unity as in GR, and can depend on the masses. 
Zhao \etal study a sample of double-image lenses from the CASTLES catalog~\cite{MunozEtAl1999} by modelling the lense as a point mass and with the Hernquist profile.
 Chen and Zhao~\cite{ChenZhao2006} and Chen~\cite{Chen2007} calculated the probability of two images occuring as a function of their separation.

Feix, Fedeli and Bartelmann analyzed the effects of asymmetric systems on gravitational lensing 
in the non-relativistic approximation of TeVeS~\cite{FeixFedeliBartelmann2007}. 
They used a Laurant expansion of the free function
and found a strong dependence of the lensing properties on the extend of the lense along the line-of-sight.
They found that each of their simulated TeVeS convergence maps had a strong resemblance with the dominant baryonic component. As a consequence they
showed (in accordance with~\cite{AngusFamaeyZhao2006}) that TeVeS cannot explain the weak lensing map of the "Bullet Cluster"~\cite{CloweEtAl2006} without an additional dark component.

Xu \etal studied the effect on large filaments on gravitational lensing in MOND and in the non-relativistic approximation of TeVeS~\cite{XuEtAl2007}.
They found that in contrast to the case of General Relativity, even if the projected matter density is zero one still gets image distorsion and magnification effects.
 They conclude that since galaxies and galaxy clusters reside in
such filaments or are projected on such structures, it complicates the interpretation of the weak lensing shear map in TeVeS. 
Thus, as they argue, filamentary structures might contribute in a significant
and complex fashion in the context of TeVeS in cases such as the "Bullet Cluster"~\cite{CloweEtAl2006} and "Cosmic wreck train" Abel520~\cite{MahdaviEtAl2007}. 

Shan, Feix, Famaey and Zhao~\cite{ShanEtAl2008} fitted fifteen double-image lenses from the CASTLES catalog using the quasistatic non-relativistic approximation of 
TeVeS and the same free function as in~\cite{ZhaoEtAl2006}. They find good fits for ten double-image lenses, however, the remaining five lenses
do not provide a resonable fit. They note that all of those five lenses are residing in or close to groups or clusters of galaxies. Since
lensing in TeVeS and more generally MOND is much more sensitive to the three-dimensional distribution of the lens and of the environment 
than in General Relativity~\cite{XuEtAl2007}, non-linear effects could be important.

Chiu, Tian and Ko~\cite{ChiuTianKo2008} performed a study of ten double-image lenses from the CASTLES catalog and twenty two lenses from the SLACS catalog.
They find good agreement of TeVeS with lensing.

Finally Mavromatos, Sakellariadou and Yusaf~\cite{MavromatosSakellariadouYusaf2009} performed the first lensing calculation in TeVeS
which explicitly involves solving the full relativistic equations with the vector field. This was done under the assumption that 
the vector field is aligned with the time direction and has no spatial component. They used lenses from the CASTLES catalog  and
 the specific choice of free function of Angus, Famaey and Zhao~\cite{AngusFamaeyZhao2006} parameterized by a single parameter.
They found that for this free function, the choice of the parameter which leads to acceptable lensing without Dark Matter leads to
very bad fits for galactic rotation curves, while the choice which gives good galactic rotation curves cannot explain lensing without additional
dark matter. It may be, however, that having a spatial component in the vector field, having a different vector field action, or using a different free function
could provide for the right lensing without compromising the galactic rotation curves.

\subsection{Superluminality, causality and gravi-Cherenkov radiation}
As shown by Elliott, Moore and Stoica~\cite{ElliottMooreStoica2005}, the Einstein-{\AE}ther theory
would lead to gravi-Cherenkov radiation, unless the speed of the spin-0 and spin-1 modes of the vector field is
superluminal. 
If this was also true in TeVeS it would be potential trouble. Furthermore how do we reconcile this statement with
the requirement of the absence of closed signal curves in the physical frame?

TeVeS could provide a solution, since the production of gravi-Cherenkov radiation should be evaluated in the
diagonal frame where the three gravitational fields decouple. It is in that frame that the speeds of all non-tensor modes
should be superluminal. However it might still be that the speeds in the physical frame are subluminal.

Bruneton~\cite{Bruneton2006} analyzed the issue of sub/superluminality in theories such as TeVeS by studying the initial value problem.
He found that superluminality need not create problems for the initial value formulation. As he showed, however, in the MOND limit $\mu\rightarrow 0$ the 
scalar field becomes non-propagating. Thus if the MOND limit is exact, there can be a singular surface surrounding each galaxy on which the scalar field does not propagate.

\subsection{Time travel of gravitational waves}
Gravitational waves in TeVeS have a different light-cone than electromagnetic waves or other massless particles due to the disformal transformation.
The speed of gravity is thus expected to be different than the speed of light by factors of $e^{-2\phi_c}$.

Kahya and Woodard~\cite{KahyaWoodard2007} (see also~\cite{Kahya2008}) have used this difference to propose a test for TeVeS and other theories with the same feature.
They propose to look at the time of arrival of gravitational waves and of neutrinos from distant supernovae.
This was further studied by Desai, Kahya and Woodard in ~\cite{DesaiKahyaWoodard2008}.

\subsection{Binary pulsars}
An important test on TeVeS is the timing of binary pulsars. Binary pulsars have provided strong tests for General Relativity. Any other theory which
aims to replace General Relativity should be able to obey the Binary Pulsar constraints. So far TeVeS has not been tested with Binary Pulsars 
which remain an open problem.

\section{How TeVeS was constructed}
\label{sec_construction}
Now that TeVeS theory has been described, we proceed to analyse its features. 

\subsection{Scalar field}
The action for the scalar field traces its roots to the Aquadradic Lagrangian theory of Bekenstein and Milgrom~\cite{BekensteinMilgrom1984},
named AQUAL. AQUAL is a casting of MOND into a proper, consistent, non-relativistic gravitational theory which effectively gives
back the MOND law.
Let us describe the transition from MOND to AQUAL in such a  way as to prepare the discussion for relativistic MOND.

\subsubsection{AQUAL}
A generic non--relativistic gravitational theory can be built as follows. We define two potentials, namely a 
Poisson potential $\Phi_P$ such that it obeys the Poisson equation, and a Newtonian potential $\Phi_N$ which
obeys Newton's 2nd law of motion. More explicitly we have 
\begin{equation}
  \nabla^2 \Phi_P = 4\pi G_N \rho,
 \label{eq:Poisson}
\end{equation}
where $G_N$ is Newton's constant and $\rho$ is the total matter density, and
\begin{equation}
  \vec{a} = - \nabla \Phi_N
\label{eq:NewtonII}
\end{equation}
where $\vec{a}$ is the acceleration of a test body due to the Newtonian potential.
To complete the theory, a relation between $\Phi_N$ and $\Phi_P$ must be given.
In the simplest case, Newtonian gravity, the two potentials are equal : $\Phi_N = \Phi_P$.

Milgrom breaks the equality of the two potentials, by modifying Newton's 2nd law of motion (\ref{eq:NewtonII}) into
\begin{equation}
 \mu_m(|\vec{a}|/a_0) \vec{a} = -\nabla \Phi_P
\label{eq:MOND_NII}
\end{equation}
where $\mu_m(x)\rightarrow 1$ as $x\gg1$ where one recovers the Newtonian limit, and $\mu_m(x)\rightarrow x$ as $x\ll1$ which is the ultra MOND limit.
In the ultra MOND limit we then have $|\vec{a}|^2 = a_0 |\nabla \Phi_P|$, which in the spherically symmetric case (with $\Phi_P \propto 1/r$ and 
$v^2\propto |\vec{a}| r$) gives constant velocity curves. The function $\mu_m$ is left free with only the two limits completely specified.

We now pass to AQUAL.  We see that in the MOND case, the two potentials are implicitely related through
\begin{equation}
\mu_m(|\nabla \Phi_N|/a_0) \nabla \Phi_N = \nabla \Phi_P
\end{equation}
Specification of AQUAL now becomes trivial. One completely eliminates any reference to the Poisson potential and writes 
the gravitational equations as
\begin{equation}
  \vec{\nabla} \cdot \left[ \mu_m(| \vec{\nabla} \Phi_N|/a_0)  \vec{\nabla} \Phi_N\right] = 4\pi G_N \rho
\end{equation}
which together with (\ref{eq:NewtonII}) fully specify the theory.
This equation is derivable from a Lagrangian with aquadratic kinetic term. The theory considered this way satisfies
all conservation laws like conservation of energy, momentum and angular momentum.

\subsubsection{Relativistic AQUAL}
In a relativistic theory, the gravitational potential is effectively defined as a small fluctuation of the
gravitational metric, through the weak field limit. This suggests that the relativistic analog of AQUAL should be a bimetric theory.
In other words, the relativistic analog of the Poisson potential $\Phi_P$ should be a metric $\metE_{ab}$ and the analog of the Newtonian
potential should be a metric $\metM_{ab}$. Furthermore, Newton's 2nd law is contained in the weak-field limit of the geodesic equation for
$\metM_{ab}$. It is therefore clear that $\metM_{ab}$ should be the universally coupled metric. On the other hand, the Poisson equation 
should be the weak-field limit of the Einstein equations for $\metE_{ab}$. 
We thus start from the prototype action $S = S_{\metE} + S_m$, where $S_{\metE}$ is the Einstein-Hilbert action (\ref{eq:S_EH}) for $\metE_{ab}$ and
$S_m$ is the matter action which only contains $\metM_{ab}$. To complete the theory we must specify a relation between $\metE_{ab}$ and $\metM_{ab}$.

Bekenstein and Milgrom~\cite{BekensteinMilgrom1984} introduce a scalar field $\phi$. The action for $\metE_{ab}$ and $\phi$ is given as
\begin{equation}
S = \frac{1}{16\pi G} \int d^4x \sqrt{-\metE} \tilde{R} - \frac{a_0^2(1+\beta_0)^2\beta_0}{8\pi G} \int d^4x \sqrt{-\metE} f(X)
\end{equation}
where $a_0$ is Milgrom's constant, $\beta_0$ is another constant and $f(X)$ is a free function. The variable $X$ is given 
by $X = \frac{\metE^{ab}\nabla_a \phi \nabla_b \phi}{a_0^2(1+\beta_0)^2}$
They complete the theory by imposing a conformal relation between $\metE_{ab}$ and $g_{ab}$ as $g_{ab} = e^{2\phi}\metE_{ab}$.
Notice that this theory has virtually all of TeVeS features except the inclusion of the vector field. Its quasistatic limit is
effectively the same as for TeVeS and reproduces the MOND phenomenology. Bekenstein and Milgrom showed, however, that the fluctuations of $\phi$ propagate 
acausally when viewed with the universally coupled metric $g_{ab}$. This is a general consequence of $f(X)$ having a MOND limit.

\subsubsection{Phase-coupling gravitation}
In an attempt to remove the acausal propagation in the relativistic AQUAL theory, Bekenstein invented the 
Phase-coupling gravitational theory~\cite{Bekenstein1988}. In this theory one replaces the real scalar field above by a complex scalar field $\chi$
which has a conventional action with respect to the metric $\metE_{ab}$, with a canonical kinetic term and a potential. The real scalar field $\phi$
is identified with the phase of $\chi$ as $\chi = |\chi| e^{i\phi}$ and is still assumed to relate the two metrics by a conformal transformation
 $g_{ab} = e^{2\eta_{p}\phi}\metE_{ab}$ where $\eta_{p}$ is a parameter. In the non-relativistic limit one finds an effective AQUAL equation for $\phi$
due to the depencence of the amplitude $|\chi|$ on $\grad\phi$ through the amplitude's field equation. All fluctuations are causally propagated in this
theory,  however, for the parameter values required by MOND phenomenology, this theory violates solar system constraints.

\subsubsection{Disformal relativistic AQUAL}
There is one other problem associated with all of the above theories, that is the bending of light. It is straightforward to show that
the bending of light associated with those theories will not be enough to account for observations without dark matter. The problem lies in the 
fact that the conformal relation does not change the lightcone structure.

Based on Bekenstein's investigation of the relation between physical and gravitational geometry~\cite{Bekenstein1993}, Bekenstein and Sanders
considered a disformal transformation of the form $g_{ab} = e^{2\phi}[ A(I) \metE_{ab} + \ell_B^2 B(I) \nabla_a \phi \nabla_b \phi]$ where $A$ and $B$ are two
functions of the invariant $I = g^{ab} \nabla_a \phi \nabla_b \phi$. The presence of the additive tensor $\nabla_a\phi\nabla_b\phi$ in the 
transformation above implies that the lightcone structure is broken, i.e. the two lightcones of $g_{ab}$ and $\metE_{ab}$ do not coincide.
However, even with this generalization, the bending of light turns out to be smaller than GR based on the visible mass alone, if the theory is to be causal.
The main obstacle is the fact that in the quasistatic limit $\nabla_\mu \phi = ( 0, \grad\phi)$, i.e. it is purely spatial, which means that 
the relation between $g_{00}$ and $\metE_{00}$ is the same as  with a conformal transformation.

\subsection{The vector field}
Sanders~\cite{Sanders1997} solved the lensing problem by introducing a unit-timelike vector field $A_a$.
The idea is to change the  relation between $g_{00}$ and $\metE_{00}$ from the one based on a conformal transformation. This would require
an additive piece like the disformal transformation above, but which would remain intact in the quasistatic limit. It is clear that
a transformation of the form $g_{ab} = A(\phi) \metE_{ab} + B(\phi) A_a A_b$  does the trick. Sanders fixed the form of $A$ and $B$ based
on the requirement of invariance under global redefinition of units and the requirement that the fine-structure constant is independent of
the frame of formulation of the theory (with $g_{ab}$ or with $\metE_{ab}$). 
This leads uniquely to the disformal transformation in TeVeS (\ref{eq:metric_relation}).
He further showed that the theory reproduces the right bending of light in a quasistatic situation.

Bekenstein~\cite{Bekenstein2004a} turned the Sanders's proposal into a fully relativistic theory by providing an action for $A_a$.
He further modified the scalar field action such that the kinetic term is $ \mu (\metE^{ab} - A^a A^b)\nabla_a \phi \nabla_b \phi$ rather than
 $ \mu\metE^{ab} \nabla_a \phi \nabla_b \phi$ as in the relativistic AQUAL theory.  This final ingredient was needed to ensure that the theory is 
is causal in all situations for which $\phi>0$. Thus TeVeS was born.

\section{Other TeVeS variants and spin-offs}
\label{sec_variants}

\subsection{Sanders Biscalar-Tensor-Vector theory}
Sanders~\cite{Sanders2005} considered a TeVeS variant which resembles the Phase-coupling gravitational theory. 
In this case the field $\mu$ is considered dynamical and has a kinetic term. Like  Phase-coupling gravitation, the pair $(\mu,\phi)$ can be combined
into a complex scalar field $\mu e^{i\phi}$ which has a conventional action. The differerence from  Phase-coupling gravitation lies in the 
use of the vector field in the disformal transformation. MOND phenomenology in quasistatic situations follows but unlike Phase-coupling gravitation
it does not require parameters which are in conflict with solar system tests. The theory also gives the
right bending of light. Sanders also considers the possibility that the Milgrom's constant $a_0$ is given by the cosmological value of $\dot{\phi}$
by building the action out of the invariants $\metE^{ab}\nabla_a \phi \nabla_b \phi$ and $A^a \nabla_a \phi$.

\subsection{The generalized Einstein-Aether theory}
Zlosnik, Ferreira and Starkman~\cite{ZlosnikFerreiraStarkman2006} showed that one can eliminate the Einstein-frame metric altogether and write TeVeS theory
using the physical metric alone. In the process, the scalar field is combined with the vector field, to define a new vector field which is dynamically
timelike but not unit. Thus TeVeS is equivalent to a vector-tensor theory (see appendix \ref{app_single_frame}).

Building upon this observation, Zlosnik, Ferreira and Starkman consider the following simplification. 
They re-insert the unit-timelike constraint of the vector field into
the physical frame, thus removing one degree of freedom. They further simplify the functions multiplying the vector field kinetic term into constants,
and the resulting action becomes the Einstein-{\AE}ther theory. To recover MOND phenomenology, they depart from the original
 Einstein-{\AE}ther theory and use a function $\mathcal{F}$ to write the action as
\begin{equation}
S = \frac{1}{16\pi G} \int d^4x \sqrt{-g}\left[ R - M_{\AE}^2\mathcal{F}\left(\mathcal{K}\right) + \lambda (A^a A_a + 1)\right]  + S_m[g]
\end{equation}
where  $M_{\AE}$ is a mass scale and
\begin{equation}
\mathcal{K} =  \left[c_1 g^{ac} g^{bd} + c_2 g^{ab} g^{cd} + c_3 g^{ad} g^{bc}\right] \nabla_a A_b \nabla_c A_d
\end{equation}
thus this theory is called the generalized Einstein-{\AE}ther theory.
In the weak field quasistatic limit one  recovers MOND phenomenology, like TeVeS.

It has been shown that the theory can lead to structure formation in a similar manner to TeVeS~\cite{ZlosnikFerreiraStarkman2008}: the vector field
plays a key role to sourcing potential wells. Cosmological studies of such theories have been conducted in the framework of the original Einstein-{\AE}ther
 theory~\cite{LiMotaBarrow2007,ZuntzFerreiraZlosnik2008} 
while the confrontation of the generalized  Einstein-{\AE}ther theory with cosmological observations is under way~\cite{ZlosnikZuntz_preparation2009}.

The theory has also been confronted with solar system tests~\cite{BonvinEtAl2008}
while gravitational lensing, in particular concerning the Bullet Cluster~\cite{CloweEtAl2006},
 has been studied by Dai, Matsuo and Starkman~\cite{DaiMatsuoStarkman2008b}.

\subsection{The generalized TeVeS theory}
\label{sec_gen_teves}
A simple generalization of TeVeS theory has been considered by Skordis~\cite{Skordis2008a}. In this case one generalizes the 
vector field action in to the same form as in the Einstein-{\AE}ther theory. In particular, we have
\begin{equation}
S_A = -\frac{1}{16\pi G} \int d^4x \sqrt{-g}\left[ K^{abcd}\tilde{\nabla}_a A_b \tilde{\nabla}_c A_d + \lambda (A^a A_a + 1)\right]  
\end{equation}
where the tensor $K^{abcd}$ is given by
\begin{equation}
K^{abcd} = c_1 \metE^{ac} \metE^{bd} + c_2 \metE^{ab} \metE^{cd} + c_3 \metE^{ad} \metE^{bc} + c_4 A^a A^c \metE^{bd}
\end{equation}
The coefficients $c_i$ are related to the ones in~\cite{Skordis2008a} as  $c_1 = K_B + K_+$, $c_2 = K_0$, $c_3 = -K_B + K_+$ and $c_4 = K_A$.
This generalization preserves the MOND limit as was explicitely calculated by Contaldi, Wiseman and Withers~\cite{ContaldiWisemanWithers2008}.

Cosmology in this theory has similar features as in TeVeS~\cite{Skordis2008a}.  It is easier to define new coefficients
$K_t = K_B + K_+ - K_A$, $\kappa_d = K_+ + \frac{1}{2}K_0 $, $K_F = 1 + K_0 + \kappa_d $ and $R_K = 1 - \frac{3\kappa_d}{K_F}$.
At the background FLRW level, the vector field simply rescales the 
gravitational constant. In particular the Friedmann equation in the Bekenstein frame becomes
\begin{equation}
3 K_F \tilde{H}^2 = 8\pi G(\rho_\phi + \rho)
\end{equation}
and so one must have $K_F >0$.
Thus as for the background dynamics, only the constant $K_F$ plays  a role by simply rescaling $G$. Therefore all FLRW solutions to TeVeS
theory would also be solutions in this general version~\cite{Skordis2008a}.

At the linearized level, only three constants play a role for scalar perturbations, namely $K_t$, $K_F$ and $R_K$. For vector perturbations
all four constants are needed and for tensor perturbations only $K_F$ and $R_K$ are important.
The primordial adiabatic mode for scalar perturbations has been constructed in~\cite{Skordis2008a} for this generalized case. The constant $K_t$
plays the same role as $K$ in TeVeS by allowing for a growing mode in the vector field in order to source structure formation.
 The constants $K_F$ and $R_K$ can also play a role and can introduce interesting features in the
power spectra~\cite{Skordis_preparation_2009b}. In particular $R_K\ne 1$ can introduce a damping term into the scalar mode $\alpha$ of the vector field
equation$\alpha$. If $R_K<1$ then $\kappa_d>0$ which damps the vector field perturbations and can stop their growth. If $R_K>1$ then
$\kappa_d<0$ which introduces an instability and the vector field grows without bound.

\subsection{The Halle-Zhao construction}
Halle and  Zhao~\cite{HalleZhao2007} considered a generalization of TeVeS, the generalized Einstein-{\AE}ther theory
and Zhao's $\nu\Lambda$ theory~\cite{Zhao2007}. They propose  vector-tensor theory in a single  frame where the vector field action
is a generalized function of both the vector field kinetic terms $\nabla_a A_b$ and the scalar $A_a A^a$ representing the magnitude of
the vector field. The vector field can be dynamically timelike or by a constraint. The different theories above come out as special cases.

\section{Outlook}
\label{sec_outlook}
I have reviewed the Tensor-Vector-Scalar theory which has been proposed as a relativistic realization of Modified Newtonian Dynamics.
The theory was a product of past antecedent theories, namely the Aquadratic Lagrangian Gravity (AQUAL) and its relativistic version~\cite{BekensteinMilgrom1984},
the phase-coupling gravitation~\cite{Bekenstein1988} the disformal relativistic scalar field theory~\cite{BekensteinSanders1994} and the Sanders's stratified
vector field theory~\cite{Sanders1997}.
Since its inception~\cite{Bekenstein2004a} it has generated a generous amount of research on various aspects, starting 
with cosmology~\cite{Bekenstein2004a,HaoAkhoury2005,SkordisEtAl2006,Diaz-RiveraSamushiaRatra2006,DodelsonLiguori2006,Skordis2006,BourliotEtAl2006,Zhao2006a,Skordis2008a,FerreiraSkordisZunkel2008}, 
spherically symmetric solutions~\cite{Bekenstein2004a,Giannios2005,JinLi2006,SagiBekenstein2008,LaskySotaniGiannios2008},
gravitational collapse and stability~\cite{Seifert2007,ContaldiWisemanWithers2008}, 
solar system tests~\cite{Bekenstein2004a,Giannios2005,BekensteinMagueijo2006,Tamaki2008,Sagi_preparation_2009},
gravitational lensing~\cite{Bekenstein2004a,ChiuKoTian2005,ChenZhao2006,ZhaoEtAl2006,Zhao2006b,Chen2007,FeixFedeliBartelmann2007,XuEtAl2007,ShanEtAl2008,ChiuTianKo2008,MavromatosSakellariadouYusaf2009}
issues on superluminality~\cite{Bruneton2006}
and the travel time of gravitational waves~\cite{KahyaWoodard2007,Kahya2008,DesaiKahyaWoodard2008}.

There are many open questions and problems still remaining to be settled in TeVeS theory. Perhaps the most looming problem awaiting solution is to
reconcile TeVeS with observations of the CMB radiation. This may require inclusion of isocurvature modes, a different free function,
or a different TeVeS action. 

Another important problem is the issue of stability, in particular of spherically symmetric perturbations, and the avoidance of caustic
singularities. This may also require changing the vector field action. 

It is also of particular importance to resolve many issues surrounding gravitational
lensing. These are issues of the environmental MOND effects on the lensing predictions, such as from filaments, which can
obscure tests of TeVeS with gravitational lensing.

Finally, it would be important to resolve problems with clusters of galaxies and galaxy groups as well as the bullet-clusters. 
It particular it is vital to elucidate the role of the vector field on those scales.
Since it has been shown that the vector field plays a fundamental role in driving large scale structure formation, it should be  deemed important
on cluster scales and may be the key to solving the problems of those systems with MOND.

Further open questions include binary pulsar tests, isocurvature modes, cosmological evolution for other free functions,
 cosmic acceleration without an effective cosmological constant, solar system tests, and de Sitter black holes.

As the influx of cosmological data continues in the next few years, theories such as TeVeS should be considered further as an explanation of the missing mass
and the missing energy problems. They provide interesting and promising research directions.

{\it Acknowledgments}:
I thank Jacob Bekenstein for suggestions and comments.
Research at the Perimeter Institute is supported in part by NSERC and by the Province of Ontario through MEDT.

\section{Bibliography}
\bibliographystyle{eprint_unsrt}
\bibliography{references}

\appendix

\section{Alternative TeVeS conventions}
In the original TeVeS article~\cite{Bekenstein2004a}, Bekenstein uses a "tilde" for the matter-frame (what he calls the physical) metric, and no "tilde" for the Einstein-frame 
(what he calls geometric) metric. Furthermore the symbol $\mathcal{U}_a$ is used for the vector field rather than $A_a$. The scalar field action, and field equations
contain an auxiliary scalar field $\sigma$, related to $\mu$ by $\mu = 8\pi G \sigma^2$ (note that Bekenstein also uses a field $\mu$ which is not the same as the one in
this review). Finally the scalar field free function is given as $F(k G \sigma^2)$ related to $V(\mu)$ as
\begin{equation}
V(\mu) = \frac{1}{16\pi\ell^2} \mu^2 F = \frac{4\pi G^2}{\ell^2} \sigma^4 F
\end{equation}

\section{Alternative and classically equivalent actions for a general class of TeVeS theories}

\subsection{The aquadratic kinetic term action}
\label{app_aqual_phi}
Since $\mu$ is a function of kinetic terms of $\phi$, we can simply replace the scalar field action with
\begin{equation}
  S = -\frac{1}{16\pi G \ell_B^2}  \int d^4x \sqrt{-\metE} f(X)
\end{equation}
where
\begin{equation}
 X = \ell_B^2 (\metE^{ab} + \beta A^a A^b) \nabla_a \phi \nabla_b \phi.
\label{eq_X}
\end{equation}
This is akin to the k-essence/k-inflation actions considered elsewhere~\cite{Armendariz-PiconDamourMukhanov1999,Armendariz-PiconMukhanovSteinhardt2000}, 
but with an additional  coupling of the vector field to the scalar.
One can then simply transform between this and the action in the main part of the text via
\begin{equation}
\mu = \frac{df}{dX}
\end{equation}
and
\begin{equation}
  f = \mu X + \ell_B^2 V
\end{equation}

This action may be more intuitive, in making contact with the MOND limit in the quasistatic case.

\subsection{The single metric frame}
\label{app_single_frame}
Zlosnik, Ferreira and Starkman have shown how to write TeVeS theory completely in the universally coupled frame~\cite{ZlosnikFerreiraStarkman2006}.
This is possible by defining a new vector field $B_a = A_a$ such that $B^a = \metM^{ab}B_b$. The magnitude of this vector field with respect to
the universally coupled metric is related to the scalar field $\phi$ as
\begin{equation}
 B^2 =  \metM^{ab}B_a B_b = -  e^{-2\phi}
\end{equation}
It is thus possible to eliminate $\phi$ from the action in terms of $B^2$ and eventually $B_a$.
At the same time, it is also possible to eliminate the metric $\metE_{ab}$ in terms of $g_{ab}$ and $B_a$ as
\begin{equation}
  \metE_{ab} = - \frac{1}{B^2}\metM_{ab} + (\frac{1}{B^4} - 1)B_a B_b    
\end{equation}
and similarly for $\metE^{ab}$.
Since the above relations are algebraic, they can be used in the action. The final step consists of changing connection from
$\tilde{\nabla}_a$ (metric compatible with $\metE_{ab}$) to $\nabla_a$ (metric compatible with $g_{ab}$).

The same procedure can also be performed  for the generalized TeVeS theory~\cite{Skordis2008a} (see section \ref{sec_gen_teves}).
A very lengthly and tedious calculation gives the  action in the universally coupled frame as
\begin{equation}
S = \int d^4x \sqrt{-\metM} \left[ R  - K^{abcd} \nabla_a B_b\;\nabla_c B_d  + \frac{1}{B^2} V(\mu)\right] + S_m[g]
\end{equation}
where the tensor $K^{abcd}$ is given by
\begin{eqnarray}
K^{abcd} &=& d_1 g^{ac} g^{bd} + d_2 g^{ab} g^{cd} + d_3 g^{ad}g^{bc}  + d_4 B^a B^c g^{bd} 
\nonumber \\
&&
+ \frac{1}{2}d_5 (g^{ad} B^b B^c + g^{bc} B^a B^d )
 + d_6 g^{ac}B^b B^d 
\nonumber \\
&&
+ \frac{1}{2}d_7 (g^{ab}  B^c B^d +  g^{cd}  B^a B^b ) + d_8 B^a B^b B^cB^d
\end{eqnarray}
with the coefficients being
\begin{eqnarray}
d_1 &=&  \frac{1+c_3-c_1}{2}    B^2 
- \frac{1}{B^2} 
+\frac{1-c_{13}}{2B^6} 
\\
d_2 &=&  \frac{1}{B^2} -\frac{1+c_2}{B^6} 
\\
d_3 &=&  \frac{c_1-c_3-1}{2}  B^2 + \frac{1-c_{13}}{2B^6}
\\
d_4 &=&  \frac{c_1-c_3-1}{2} + \frac{c_4-c_1 + 1}{B^4}  +  \frac{c_{13}-1}{2B^8} 
\\
d_5 &=&  1 +  c_3 -c_1 + \frac{2(c_1-c_4-2)}{B^4} + \frac{c_{13}-1}{B^8}  
\\
d_6 &=& \frac{c_1-c_3-1}{2} 
+ \frac{3+c_4-c_1 + \mu}{B^4} 
 + \frac{c_{13}-1}{2B^8} 
\\
d_7 &=& - \frac{2}{B^4} + \frac{2(3 + c_{13} + 4c_2)}{B^8} 
\\
d_8 &=&   \frac{2-\mu}{B^6}  + \frac{ (1-\beta)\mu - 6c_{13}-16c_2 - 10 }{B^{10}}  
\end{eqnarray}
and where the constants $c_i$ appear in the vector field action (see section \ref{sec_gen_teves}).
In the case of the original TeVeS,  the above coefficients differ from  those in~\cite{ZlosnikFerreiraStarkman2006}
which are incorrect. The correct coefficients can also be found in Zlosnik's PhD thesis~\cite{Zlosnik_Thesis}.

Notice that the time-like constraint is now absent, since it has been used to eliminate the scalar field. The scalar field
is absorbed into the vector field, which now has four, rather than three independent degrees of freedom.
The fact that the vector field is timelike is now imposed dynamically, due to the presence of inverse powers of its modulus $B^a B_a$,
i.e. the vector field can never be null. It seems that this action describes two disconnected sectors, one where the vector field is
timelike and one where it is spacelike, and any transitions between them are forbidden. However the spacelike section, when expanded
around the background value $B^2 = 1$, leads to wrong-sign kinetic terms. Therefore only the timelike sector is the physical one, which is precisely TeVeS theory.

\subsection{The diagonal frame}
\label{app_diagonal}
In the diagonal frame, one performs a yet another disformal transformation to a new metric $\metS_{ab}$, as well as 
a field redefinition of $\phi$ and $A_a$. 
We shall consider a transformation of a more general version of TeVeS. The vector field action is generalized to 
an Einstein-{\AE}ther action with coefficients $c_i$ (see section \ref{sec_gen_teves}) as in~\cite{Skordis2008a}.
We also generalize the scalar field action to
\begin{equation}
S_\phi = -\frac{1}{16\pi G} \int d^4x \sqrt{-\metE}\left[ \mu (\metE^{ab} + \beta A^a A^b ) \nabla_a \phi \nabla_b\phi + V(\mu)\right]
\end{equation}
where the new parameter $\beta$ is equal to $\beta=-1$ in the original TeVeS theory,
such that $\beta<1$ to preserve the metric signature.

The new fields are defined by
\begin{eqnarray}
 \hat{g}^{ab} &=& \frac{1}{\sqrt{1-\beta}}[\tilde{g}^{ab}  + \beta A^a A^b],
\\
 \hat{g}_{ab} &=& \sqrt{1-\beta}\tilde{g}_{ab}  - \frac{\beta}{\sqrt{1-\beta}}A_a A_b,
\\
  \chi &=& \phi + \frac{1}{4}\ln (1-\beta),
\\
 \hat{A}_a &=&   (1-\beta)^{-1/4} A_a,
\\
 \hat{A}^a &=& (1-\beta)^{1/4} A^a.
\end{eqnarray}
The new vector field $\hat{A}_a$ remains unit-timelike with respect to the new metric : $\hat{A}_a \hat{A}^a  = \hat{g}^{ab} \hat{A}_a\hat{A}_b = -1$.
Notice that for $\beta=-1$, the inverse metric transformation takes the same form as the one appearing in the scalar field action in
TeVeS. Thus with this transformation,  the scalar field completely decouples from the vector field in this frame.
With respect to this metric, as the name implies, the kinetic terms of the new metric, the scalar field and the new vector field
become diagonalized.
Disregarding the scalar field, this transformation has been considered by Foster~\cite{Foster2005}, in the context of Einstein-{\AE}ther theory.

Under this transformation, the TeVeS action becomes
\begin{eqnarray}
S &=&  \frac{1}{16\pi G} \int d^4x \sqrt{-\metS} \bigg[ \hat{R}  - \KS^{abcd} \nablaS_a \AS_b\;\nablaS_c \AS_d  + \lambda (\AS^a \AS_a +1) 
\nonumber 
\\
&& 
- \mu \metS^{ab}\nabla_a \chi \nabla_b \chi - \hat{V}(\mu)\bigg] + S_m[g]
\end{eqnarray}
where the Lagrange multiplier has been appropriately rescaled, $\hat{V} = V/\sqrt{1-\beta}$ and where 
\begin{eqnarray}
\KS^{abcd} &=& \cS_1 \metS^{ac} \metS^{bd} + \cS_2 \metS^{ab} \metS^{cd} + \cS_3 \metS^{ad}\metS^{bc}  + \cS_4 \AS^a \AS^c \metS^{bd}.
\end{eqnarray}
The new coefficients $\cS_i$ are related to the old ones $c_i$ via 
\begin{eqnarray}
   \cS_1 &=& \frac{ (2-2\beta + \beta^2) c_1
+ \beta (2-\beta) c_3  - \beta^2 }{2(1-\beta)} ,
\nonumber 
\\
  \cS_2 &=&  \frac{c_2 + \beta }{1-\beta}  ,
\nonumber 
\\
\cS_3 &=& \frac{ (2-2\beta+\beta^2) c_3
+\beta (2-\beta) (c_1  - 1) }{2(1-\beta)}  ,
\nonumber 
\\
\cS_4 &=& c_4 + \frac{ \beta^2 (c_1 -1)
+\beta(2-\beta) c_3 }{2(1-\beta)}
\end{eqnarray}

Variation of the action proceeds by noting that the physical metric is now given in terms of $\metS_{ab}$ as
\begin{equation}
\metM_{ab} = e^{-2\chi} \metS_{ab} - 2\sinh(2\chi)  \AS_a \AS_b
\end{equation}
which is the same as the standard TeVeS transformation.
Note that in the case of the original TeVeS theory, for which $\beta=-1$, $\tilde{c}_2 = \tilde{c}_4 = 0$ and $\tilde{c}_1 = K = -\tilde{c}_3$ we get
\begin{eqnarray}
   \cS_1 = 2K - \frac{ 1 }{4} 
&\qquad&
  \cS_2 =  -\frac{1}{2}  
\\
\cS_3 =  -2K +\frac{ 3 }{4}
&\qquad&
\cS_4 =  K - \frac{1}{4}
\end{eqnarray}
while in terms of the linear combinations $K_t$, $\kappa_d$, $K_F$ and $R_K$ as in~\cite{Skordis2008a} (see section \ref{sec_gen_teves}) we get
\begin{eqnarray}
   K_t = K 
&\qquad&
   \kappa_d = 0
\\
 K_F = \frac{1}{2}
&\qquad&
R_K = 1
\end{eqnarray}
Thus this transformation does not introduce any damping term $\kappa_d$.


\end{document}